\documentclass{aastex62}

\newcommand{\teff}{\mbox{T$_{\rm eff}$}}
\newcommand{\logg}{\mbox{log~{\it g}}}
\newcommand{\vmicro}{\mbox{$\xi_{\rm t}$}}
\newcommand{\kmsec}{\mbox{km~s$^{\rm -1}$}}
\newcommand{\loggf}{\mbox{$\log gf$}}

\newcommand{\x}{\mbox{$\Delta_{\tiny{\mathrm{F275W,F814W}}}$}}
\newcommand{\y}{\mbox{$\Delta_{\tiny{C~\mathrm{ F275W,F336W,F438W}}}$}}

\submitjournal{ApJ}

\shorttitle{Chemical abundances in NGC\,6934} 
\shortauthors{A.\,F. Marino, et al.} 

\begin{document}

\title{METALLICITY VARIATIONS IN THE TYPE~II GLOBULAR CLUSTER NGC\,6934\footnote{Based on observations with  the NASA/ESA {\it Hubble Space Telescope}, obtained at the Space Telescope Science Institute, which is operated by AURA, Inc., under NASA contract NAS 5-26555. This paper includes data gathered with the 6.5 meter Magellan Telescopes located at Las Campanas Observatory, Chile, and Gemini Telescope at Canada-France-Hawaii Telescope. }}

\author{A.\ F.\,Marino} 
\affiliation{Research School of Astronomy \& Astrophysics, Australian National University, Canberra, ACT 2611, Australia} 
\author{D. Yong}
\affiliation{Research School of Astronomy \& Astrophysics, Australian National University, Canberra, ACT 2611, Australia}  
\author{A.\ P.\,Milone}
\affiliation{Dipartimento di Fisica e Astronomia ``Galileo Galilei'' - Univ. di Padova, Vicolo dell'Osservatorio 3, Padova, IT-35122}
\author{G.\ Piotto}
\affiliation{Dipartimento di Fisica e Astronomia ``Galileo Galilei'' - Univ. di Padova, Vicolo dell'Osservatorio 3, Padova, IT-35122}
\author{M.\, Lundquist}
\affiliation{Gemini Observatory, Northern Operations Centre, 670 North A'ohoku Place, Hilo, HI 96720, USA}
\author{L.\ R.\,Bedin}
\affiliation{Istituto Nazionale di Astrofisica - Osservatorio Astronomico di Padova, Vicolo dell'Osservatorio 5, Padova, IT-35122}
\author{A.-N.\, Chen\'e}
\affiliation{Gemini Observatory, Northern Operations Centre, 670 North A'ohoku Place, Hilo, HI 96720, USA}
\author{G.\ Da Costa}
\affiliation{Research School of Astronomy \& Astrophysics, Australian National University, Canberra, ACT 2611, Australia}  
\author{M.\ Asplund}
\affiliation{Research School of Astronomy \& Astrophysics, Australian National University, Canberra, ACT 2611, Australia}  
\author{H.\ Jerjen}
\affiliation{Research School of Astronomy \& Astrophysics, Australian National University, Canberra, ACT 2611, Australia}  

\correspondingauthor{A.\ F.\,Marino}
\email{anna.marino@anu.edu.au}

\begin{abstract}
The {\it Hubble Space Telescope} photometric survey of Galactic
globular clusters (GCs) has revealed a peculiar ``chromosome map'' for
NGC\,6934. Besides a typical sequence, similar to that observed in
Type~I GCs, NGC\,6934 displays additional stars on the red side,
analogous to the $anomalous$, Type~II GCs, as defined in our previous
work. 
We present a chemical abundance analysis of four red giants in this
GC. Two stars are located on the chromosome map sequence common to all
GCs, and another two on the additional sequence. 
We find: {\it (i)} star-to-star Fe variations, with the two
$anomalous$ stars being enriched by $\sim$0.2~dex. Due to our
small-size sample, this difference is at the $\sim$2.5~$\sigma$ level; 
{\it (ii)} no evidence for variations in the $slow$ neutron-capture
abundances over Fe, at odds with what is often observed in $anomalous$
Type~II GCs, e.g.\,M\,22 and $\omega$~Centauri; 
{\it (iii)} no large variations in light elements C, O and Na,
compatible with the targets location on the lower part of the
chromosome map where such variations are not expected. 
Since the analyzed stars are homogeneous in light elements, the only
way to reproduce the photometric splits on the 
sub-giant (SGB) and the red-giant (RGB) branches is to assume that
red-RGB/faint-SGB stars are enhanced in [Fe/H] by $\sim$0.2. This fact
corroborates the spectroscopic evidence of a metallicity variation in
NGC\,6934. 
The observed chemical pattern resembles only partially the other
Type~II GCs, suggesting that NGC\,6934 might belong either to a third
class of GCs, or be a link between $normal$ Type~I and $anomalous$
Type~II GCs. 
\end{abstract}


\section{Introduction}\label{sec:intro}

In the last few years there have been an increasing number of
observations that indicate a sizeable sub-sample of the Milky Way
globular clusters (GCs) host 
stellar populations with different metallicities (here intended as
primarily Fe). 
These variations 
are in addition to the
typical star-to-star variations
observed for elements involved in hot-H burning processes, e.g. C, N, O, Na, and
sometimes Mg, and Al. 

The {\it Hubble Space Telescope} ($HST$) UV-Legacy of Galactic GCs (Piotto
et al.\,2015) has recently
revealed that the multiple stellar populations phenomenon in GCs is
best described by the so-called ``chromosome map'', a diagram
constructed from a combination of $HST$ filters whose $x$ (\x) and $y$ (\y) axes
are mostly sensitive to helium and nitrogen, respectively (Milone et
al.\,2015, 2017). 
These diagrams were given their name because they represent a imprint
of the formation processes that occurred in GCs which produced
what we observe today as multiple stellar populations. 
The morphology of the chromosome maps varies from cluster to
cluster, displaying different extensions and numbers of
overdensities, corresponding to different stellar populations.

Despite some degree of variation between the maps, 
common features have been observed (Renzini et al.\,2015; Milone et
al.\,2017). First, {stars in all the GCs display a 
common pattern
in their chromosome map extending from relatively
low \y\ and high \x\ values, to lower \x\
and higher \y. Second, a first stellar generation (1G), defined by the
population of stars located at the lower \y\ values of the map, has
been identified in all GCs investigated so far. 
Stars associated with the second generation (2G) are found at
  higher \y\, defining a sequence in both \x\ and \y, that extends
  from lower-\y/higher-\x, towards different degrees of enhanced \y\
  values. As these feauture have been observed as a general behavior
  in Milky Way GCs, the GCs showing exclusively these features have
  been named Type~I GCs (Milone et al.\,2017).

However, ten of the 57 GCs in the {\it HST} study have photometric
peculiarities (Milone et al.\,2017).  
These objects, named Type~II GCs, in addition to the common 1G and 2G
stars observed in Type~I GCs, display additional sequences in the red
side of their chromosome map. 
These red stars usually follow the same pattern observed for the 1G
and 2G stars in Type-I GCs, i.e. a distribution from high-\x/low-\y\
to lower-\x/higher-\y. However, the lowest \y\ values observed for the
red stars changes from cluster to cluster. 

All the Type~II GCs for which chemical abundances are available from
spectroscopy have been classified as $anomalous$ GCs, i.e. as 
GCs exhibiting internal variations not just in the light elements
involved in hot-H burning processes, but also in the bulk metallicity
and in heavier elements, specifically in elements produced via {\it
  slow} neutron-capture reactions ([$s$-elements/Fe]).  
Type~II GCs include NGC\,1851 (Yong \& Grundahl\,2008), M\,22
(Marino et al.\,2009, 2011; Lee\,2016), NGC\,5286 (Marino et al.\,2015), M\,2
(Yong et al.\,2014), and,
the most extreme case, $\omega$~Centauri (e.g.\, Norris \&
Da Costa 1995; Johnson \& Pilachowski\,2010; Marino et al.\,2011).
In all cases variations in
[$s$-elements/Fe] are positively correlated with Fe .
Further, in NGC\,1851 (Yong et al.\,2009, 2015), M\,22 (Marino et al.\.2011) and
$\omega$~Centauri (Marino et al.\,2012) a difference
in the overall C+N+O abundance has been found, with the stars enriched in
[$s$-elements/Fe] also having higher [C+N+O/Fe] 
abundances. As stellar populations with different C+N+O are expected to
have distinct luminosities on the sub-giant branch (SGB), these chemical
variations are considered to play an important role in generating the split SGBs
observed in these three GCs (Milone et al.\,2008; Cassisi et
al.\,2008; Marino et al.\,2012a,b).

The complex chemical pattern of $anomalous$ clusters and its
resemblance on a less extreme scale to $\omega$~Centauri have raised the
intriguing idea that these GCs could have originated in
extra-Galactic environments (Bekki \& Freeman\,2003). Specifically,
these GCs could be the  
surviving nuclei of now-disrupted dwarf galaxies, similar to what
has been
proposed to explain the chemical variations in $\omega$~Centauri. Such
a dwarf galaxy environment could explain 
the capability of these GCs to support more extended star formation histories
than typical GCs which show variations in light elements only.
 
In this context, M\,54 stands out as an intriguing case (Carretta et
al.\,2010). It is the nuclear star 
cluster of the Sagittarius dwarf galaxy (Bellazzini et al.\,2008) and
certainly formed and 
evolved in an extra-Galactic environment.  
The photometric and spectroscopic similarities between M\,54 and the
other $anomalous$ GCs, makes it tempting to speculate that the latter
are indeed remnants of dwarf galaxies cannibalized by the Milky Way. 

To date heavy elements variations have been observed in 11 Galactic GCs. 
Most surprisingly, the $HST$ UV Survey has revealed that a relatively
large fraction of the total sample of the 57 observed clusters,
$\sim$18\%, belongs to the 
class of Type~II GCs. We don't know yet if all Type~II GCs
can also be classified as $anomalous$ from their chemistry, but all
the $anomalous$ GCs behave as Type~II, based on their chromosome maps. 
The idea that such a large fraction of GCs ($\sim$18\%) could be
chemically $anomalous$ would mean that many GCs in the Milky Way
had a deeper potential well with which to retain high velocity
supernovae ejecta required to explain metallicity variations, and
possibly be dwarf galaxy remnants. 

In the present study we continue our investigation of Type~II
GC properties.
For that purpose, we focus on the chemical abundances of a poorly
studied GC, NGC\,6934, which exhibits an unusual chromosome map.
This metal-intermediate cluster ([Fe/H]$=-$1.47, Harris 1996, as
updated in 2010) has been classified as a Type~II GC since it hosts a split
SGB visible in the optical filters which is connected with a red red-giant
branch (RGB) that is
absent in typical Milky Way GCs (Milone et al.\,2017). 
On the other hand, the position of these red RGB stars is slightly
different from what is observed in $anomalous$ GCs, as they lie at lower
average \y\ values.
Proper motions suggest that the {\it red} sequence stars on the
chromosome map are cluster members, but they have never been
investigated for chemical abundances.  
To better understand the chemical properties of multiple stellar populations
in NGC\,6934 we have selected four targets, two in each of the two
sequences identified in its chromosome map.
 
The outline of the paper is as follows: Section~\ref{sec:data} is a
description of the spectroscopic and photometric data-sets; the choice
of the adopted atmospheric parameters is discussed in
Section~\ref{sec:atm}, while our chemical abundance analysis is outlined in
Section~\ref{sec:abundances}. Section~\ref{sec:abb} describes the
results, which are summarized and discussed in Section~\ref{sec:conclusions}.

\section{Data}\label{sec:data}

\subsection{The photometric dataset: the chromosome map of NGC\,6934 \label{sec:phot_data}} 

The photometric data used in this study come from the $HST$ UV Legacy
Survey which investigated multiple stellar populations in GCs (GO-13297,
Piotto et al.\,2015).
Details on the images analyzed and on the data reduction can be found in
Piotto et al.\,(2015) and Milone et al.\,(2017).
Photometry has been corrected for differential reddening effects,
which are very small, $\Delta(E(B-V))\lesssim$0.004~mag, for NGC\,6934.

Milone et al.\,(2017) analyzed the chromosome map
of 57 GCs, including NGC\,6934, and noticed some peculiarities (see
their Fig.~4) for this cluster with respect to typical GCs maps.  
The chromosome map of NGC\,6934 red giants is reproduced on the left
panel of Fig.~\ref{fig:cmdtarget}, 
with the dashed line separating 2G from 1G stars as defined in Milone
et al.\,(2017). 
In this parameter space we can clearly distinguish:
{\it (i)} the presence of 1G and 2G stars, as the stars located
  below and above the dashed line, respectively; and {\it (ii)}
 two distinct patterns
of stars represented with gray and red dots, respectively.
The presence of the  separate distribution of red stars
on the chromosome map is a distinctive feature of the Type~II GCs.

Another distinctive feature of Type~II GCs is the presence of a split
SGB in both optical and UV color-magnitude diagrams (CMDs). As shown
in the right panel of 
Fig.~\ref{fig:cmdtarget}, the faint SGB of NGC\,6934 is clearly
connected with the red 
RGB in the $m_{\rm F336W}$ vs.\,$m_{\rm F336W}-m_{\rm F814W}$ CMD. 
Red stars on the chromosome map have been selected as the stars
defining the red RGB on the CMD, represented in the right panel of
Fig.~\ref{fig:cmdtarget}.  

From its optical and ultraviolet CMDs and from its chromosome map,
NGC\,6934 has been classified as a Type II GC, though, 
as discussed in Section~\ref{sec:intro},
the comparison of the chromosome map of
NGC\,6934 with that observed in {\it anomalous} GCs suggests that 
the majority of {\it red} stars in {\it anomalous} GCs have higher
values on the \y\ axis.  
This difference in the distribution of the stars can be easily seen in
Fig.~\ref{fig:n6934_m22_n6752} where we represent the chromosome map of
NGC\,6934 (left panel) in comparison with that of the {\it anomalous} GC M\,22
(middle panel). For comparison purposes, we also display the
chromosome map of the Type~I GC NGC\,6752 (right panel), which
exhibits the presence of just 1G and 2G stars.
In all the three GCs shown in Fig.~\ref{fig:n6934_m22_n6752}, the
dashed line represents the separation between 1G and 2G stars, as
defined in Milone et al.\,(2017). 
It is clear that while in M\,22 all the red stars have \y\ values
above the 1G-2G separation line, some red stars in NGC\,6934 lie below
this line.

To avoid confusion, we repeat here the terminology that will be used
throughout the paper. 
Firstly, according to their chemical compositions, GCs are classified into
{\it normal} and {\it anomalous} (e.g. Marino et al.\,2015). 
Following this classification, in this paper the terms {\it normal} and {\it
  anomalous} will refer to the chemical pattern observed in GCs from
spectroscopy: 
\begin{itemize}
\item{{\it normal} are the monometallic systems with the typical
    chemical patterns 
    observed in GCs, e.g.\ the (anti-) correlations in light elements;}
\item{{\it anomalous} will indicate GCs with unusual chemical
    pattern in heavy elements, e.g.\ variations in the
    [$s$-element/Fe] ratios and/or [Fe/H];}
\end{itemize}
Secondly, according to their photometric pattern (Milone et
al.\,2017), GCs will be identified as:  
\begin{itemize}
\item{Type~I, which will be used for GCs with single SGBs in CMDs from
photometry with optical filters. These objects display the typical
chromosome map characterised by a {\it single} sequence;}
\item{Type~II, which designates those GCs with a split SGB in
    optical filter CMDs. These clusters exhibit a blue and red RGB in
    the $m_{\rm F336W}$ vs.\,$m_{\rm F336W}-m_{\rm F814W}$ CMD. 
    Stars on red RGB also define additional red sequences on
    the chromosome map. }  
\end{itemize}
The synergy between photometry and spectroscopy reveals that:
{\it (i)} {\it normal} GCs  exhibit the CMDs and the chromosome maps
of Type I GCs; 
{\it (ii)} all the {\it anomalous} GCs analyzed so far are Type~II and the
 stars with enhanced Fe and/or [$s$-element/Fe] populate the red RGB. 
In the following we will also refer to individual stars in Type~II GCs as:
\begin{itemize}
\item{blue- or {\it normal} RGB stars if they define the blue RGB and
    fall in the typical
    chromosome map sequence, as observed in all GCs;}
\item{red- or {\it anomalous} RGB stars if they are located on the additional
  red sequences, observed in Type~II GCs only;}
\item{in this paper the targets on the {\it normal} and {\it
      anomalous} RGBs will be designated n1, n2 and a1, a2, respectively.}
\end{itemize}

\subsection{The spectroscopic dataset}\label{sec:spec_data}

Our spectroscopic data have been acquired using the Gemini Remote Access
to CFHT ESPaDOnS (Donati\,2003) 
Spectrograph (GRACES; Chen\'e et al.\,2014) through the program
GN-2015B-Q-81, and with the Magellan Inamori Kyocera
Echelle (MIKE, Bernstein et al.\,2003) spectrograph on the
Magellan-Clay 6.5m telescope. 
The targets were four RGB stars with $V \sim$15~mag
selected from the chromosome map described in Section~\ref{sec:phot_data}. 
Their location in the chromosome map
and in the $m_{\rm F336W}$-$(m_{\rm F336W}-m_{\rm F814W})$ CMD of
NGC\,6934 are shown in Fig.~\ref{fig:cmdtarget}.
We have chosen two giants lying on the 
{\it normal} RGB and two on the {\it anomalous}
RGB, represented with blue filled circles and red stars,
respectively. 
As our primary goal is to study the heavy element pattern, 
we note here that our spectroscopic sample are biased
towards relatively low values of \y. This choice was made to avoid
contamination from
stars characterised by substantially different abundances of nitrogen
and/or other light elements (Milone et al.\,2017).
Furthermore, our selected targets are among the brightest 1G stars in
order to maximize the S/N of the spectra. 
The observing log is given in Tab.~\ref{tab:data}.

High resolution ($R\sim 40,000$) optical spectra were obtained with
GRACES in August and December 2015 using
the target$+$sky 2-fibre spectroscopic mode. 
Briefly, light from the Gemini North telescope is
fed to the Echelle SpectroPolarimetric Device for the Observation of
Stars (ESPaDOnS) at the Canada-France-Hawaii Telescope (CFHT) via two
270~m long optical fibres which have $\sim$8 per cent throughput (see Chen\'e et
al.\,2014).
Basic data reduction was performed using the OPERA pipeline (Martioli et
al.\,2012, Malo et al., in preparation).
Two of the four targets observed with GRACES, were subsequently re-observed
with MIKE with the aim of gathering independent analysis from
a different data set.
Specifically, we gathered MIKE spectra for one target on
the {\it normal} sequence, and one on the {\it anomalous} sequence,
chosen to have very similar
effective temperatures based on the analysis of the GRACES data.
The MIKE spectra were taken in 2017 May, using a 0.70\arcsec\ slit
which gives $R\sim 35,000$ in the blue and  $R\sim 30,000$ in the red,
respectively.
Data reduction involving bias-subtraction, flat-field correction,
wavelength-calibration, and sky-subtraction, has been done by using the
dedicated CarPy pipeline\footnote{See {\sf
    http://code.obs.carnegiescience.edu/mike}} (Kelson et al.\,2000;
Kelson\,2003). 

Both for the GRACES and MIKE spectra, co-addition of the individual
exposures, continuum normalisation, radial velocity (RV) determination
and correction to laboratory wavelength was performed using IRAF routines.
Spectral coverage for the GRACES spectra is $\sim$4070-10020~\AA,
while the MIKE spectra cover the spectral region from
$\sim$3350~\AA\ to $\sim$9420~\AA. The S/N ratio of our MIKE spectra
is $\sim$120@6000~\AA, while the GRACES spectra have a lower S/N, between
40 and 70@6000~\AA.
The chemical abundance analysis was performed by neglecting the bluest
part of the spectra because of the low S/N. 
Specifically, we have analysed the spectral regions with $\lambda
  \gtrsim$4200~\AA\ for MIKE, and $\lambda
  \gtrsim$4500~\AA\ for GRACES.

Radial velocities were derived using the IRAF@FXCOR task, which
cross-correlates the object spectrum with a template.  
For the template we used a synthetic spectrum generated with the
latest version of MOOG\footnote{{\sf
    http://www.as.utexas.edu/~chris/moog.html}.} (Sneden\,1973). 
This spectrum was computed with a model stellar atmosphere interpolated 
from the Kurucz (1992)\nocite{kur92} grid\footnote{{\sf
    http://kurucz.harvard.edu/grids.html}}, adopting parameters  
(effective temperature (\teff), surface gravity (\logg),
microturbulent velocity (\vmicro), and overall metallicity ([A/H])) of
4600~K, 2.5, 2.0~\kmsec, and $-$1.45~dex, respectively. 
Observed RVs were then corrected to the heliocentric system. 

Heliocentric RVs were used as a proof of cluster membership for
our targets. We measure an 
average RV for the four stars of $<$RV$> = -405.5\pm2.9$~\kmsec\
(rms$=5.0$~\kmsec), 
which is consistent with the Harris tabulated value of
$-411.4\pm1.6$~\kmsec\ (Harris\,2010).
Such extreme negative RVs are clearly distinct from the RVs of
typical field stars. 
Hence, based on both RVs and proper motions, 
we can be confident of the membership for all four targets.
Coordinates and RVs, both observed (RV) and corrected to the heliocentric system
(RV$_{\rm helio}$), for all four stars are listed in
Tab.~\ref{tab:data}.   

\section{Model atmospheres}\label{sec:atm}

The relatively high resolution and the large spectral coverage of our
spectra allowed us to have a fully-spectroscopic estimate of the
stellar parameters, \teff, \logg, [A/H] and \vmicro. 
Hence, we determined \teff\ by imposing the excitation potential (E.P.)
equilibrium of the Fe\,{\sc i} lines and gravity with the ionization
equilibrium between Fe\,{\sc i} and Fe\,{\sc ii} lines. Note that for
\logg\ we impose Fe\,{\sc ii} abundances that are
0.05-0.07~dex higher than the Fe\,{\sc i} ones to adjust for non-local
thermodynamic equilibrium (non-LTE) effects (Bergemann et al.\,2012; Lind,
Bergemann \& Asplund\,2012). For this analysis, \vmicro\ was set to
minimize any dependence of Fe\,{\sc i} abundances as a function of EW. 
For the two stars observed with both GRACES and MIKE, we derived the
atmospheric parameters separately for each spectrum. 

As an independent test of our results, we also derived atmospheric 
parameters from our $HST$ photometry (see Section~\ref{sec:phot_data}). 
The $m_{\mathrm{F606W}}$ and $m_{\mathrm{F814W}}$ mag have been
converted to $V$ and $I$ mag (Anderson et al.\,2008), 
which we then used to estimate temperatures
from the Alonso et al.\,(1999) color-temperature calibrations.
The use of the $(V-I)$ color for this purpose is justified by the fact that 
it is mostly insensitive to variations in light elements.
Surface gravities were then obtained from the apparent $V$ magnitudes,
the photometric \teff, bolometric corrections from Alonso et
al.\,(1999), apparent distance modulus of $(m-M)_{V}=$16.28 (Harris
2010), and masses of 0.80~$M_{\odot}$. 
Once \teff\ and \logg\ have been fixed from photometry, we derived
\vmicro\ from the Fe\,{\sc i} lines as explained above.
The atmospheric parameters obtained from spectroscopy and
photometry are listed in Tab.~\ref{tab:atm}. 

By comparing the spectroscopic stellar parameters obtained from the GRACES and
MIKE spectra for the two stars observed with both instruments we
notice that: 
{\it (i)} GRACES \teff\ are only marginally higher; 
{\it (ii)} \logg\ values agree within $\lesssim$0.20~dex; 
{\it (iii)} GRACES metallicities are systematically higher by
$\sim$0.20~dex.
By adopting exactly the same atmospheric parameters from
  photometry (right columns in Tab.~\ref{tab:atm}) for the MIKE
  and GRACES spectra, we have a mean difference in [Fe/H] of 0.135~dex
  between stars observed with both the instruments, with higher values
  for GRACES\@. We have verified that this difference is introduced by the
  systematically higher EWs, by $\sim$6~m\AA, obtained from the GRACES spectra. 
We note here that the MIKE Fe abundances are in better accord with the
tabulated Harris value ([Fe/H]=$-$1.47~dex; Harris 2010). 
As abundance offsets are not unexpected when using spectra from different
spectrographs, in the following we will consider results from different
instruments separately, and average them appropriately only when necessary.

We adopt as our primary set of atmospheric parameters
the values obtained from the MIKE spectra, as they have higher S/N
and a larger number of measured Fe lines. For GRACES, due to the low
number of measurable lines, in particular Fe~{\sc ii}, we prefer to
adopt the parameters based on photometry as our primary set.  
In the following, when appropriate, we will use both the sets of
 parameters to check our results.

In order to have an estimate of the internal errors associated with
our atmospheric parameters, we have compared our
\teff/\logg\ values from the spectral lines with those derived from
the $(V-I)$ $HST$ colors.  
We obtain: $\Delta$\teff=\teff$_{\rm {Fe~lines}}
-$\teff$_{(V-I)}=+126\pm21$~K (rms=48~K),   
and  $\Delta$\logg=\logg$_{\rm {Fe~lines}}
-$\logg$_{(V-I)}=+0.06\pm0.06$ (rms=0.14). 
This comparison suggests that the spectroscopic \teff/\logg\ scales
are systematically higher, but 
the internal errors in these parameters
are smaller, comparable with the rms of the average
differences, e.g. about 50~K and 0.15~dex, in temperature and gravity,
respectively.  
Furthermore, in our chemical abundance analysis we adopt typical internal
uncertainties of 0.20~\kmsec\ for \vmicro\ and 0.10~dex for metallicity.

\section{Chemical abundances analysis}\label{sec:abundances}

Chemical abundances were derived from a LTE analysis by using the
spectral analysis code MOOG (Sneden\,1973), and 
the alpha-enhanced Kurucz model atmospheres of
Castelli \& Kurucz (2004), whose parameters have been obtained as
described in Section~\ref{sec:atm}.

A list of our analyzed spectral lines, with their associated
equivalent widths (EWs), excitation
potentials (EPs) and total oscillator strengths (log~$gf$), is provided
in Tab.~\ref{tab:lines}. The chemical abundances for all the
elements, with the exception of 
those discussed below, have been inferred from an EW-based analysis.
We now comment on some of the transitions that we used.

{\it Proton-capture elements:}
We have derived abundance ratios with respect to Fe for C, O, Na, Al
and Mg.
Carbon has been inferred from spectral synthesis of the CH G-band
$(A^{2}\Delta - X^{2}\Pi)$ features near 4312 and 4323~\AA.
The S/N of the GRACES spectra at these wavelengths was too low to derive
abundances, so C has been measured only from MIKE spectra.   
Oxygen abundances were inferred from the synthesis of the forbidden
[O\,{\sc i}] lines at 6300~\AA\ and 6363~\AA.
Although we have not derived any 3D-non-LTE correction for these
spectral lines, we notice that from the grid by Amarsi et al.\,(2016)
that they are small for metal-poor stars within 
their analysed range of atmospheric parameters
(5000$<$\teff$<$6500~K, 3.0$<$\logg$<$5.0~g/cm~s$^{-2}$). 
Our giants are, however, somewhat cooler, and have much lower
surface gravities. Further,
telluric O$_{2}$ and H$_{2}$O spectral absorptions often affect the O
line at 6300~\AA. Indeed, for our targets,  the analyzed O
transition is contaminated by O$_{2}$ lines. We have removed the telluric
features by using the software MOLECFIT\footnote{{\sf
    http://www.eso.org/sci/software/pipelines/skytools/molecfit}}
provided by ESO (Smette et al.\,2014; Kausch et al.\,2014). Nevertheless, even
with such a subtraction procedure, we caution that residual telluric
feature contamination might be of concern for the analysis of the
6300.3 [O\,{\sc i}] line. 
We determined Na abundances from the EWs of the Na\,{\sc i} doublets at
$\sim$5680~\AA, $\sim$6150~\AA\ and $\sim$8190\AA. Chemical abundances
for this species have been corrected for non-LTE effects by following
the recipes in Lind et al.\,(2011). 
Aluminum was inferred from the synthesis of the doublet at
$\sim$6667~\AA, while magnesium abundances were determined from the EWs of the
transitions at $\sim$5528, 5711~\AA. 

{\it Manganese:}
For Mn, we have synthetised the spectral lines at around
4710, 4739, 4754, 4762, 5395, 5420, 5433, 6014, 6022~\AA, by assuming 
f(${\phantom{}}^{55}$Mn)=1.00. When available, the hyperfine splitting
data have been taken from Lawler et al. (2001a,b), otherwise from the
Kurucz (2009) compendium\footnote{Available at: {\sf
    http://kurucz.harvard.edu/}}.   

{\it Copper:}
Abundances for Cu were inferred from synthesis of the Cu\,{\sc i}
lines at 5105, 5782~\AA. Both hyperfine and isotopic splitting were
included in the analysis, using the well-studied spectral line component
structure from Kurucz (2009). Solar-system isotopic fractions
were assumed in the computations: f(${\phantom{}}^{63}$Cu)=0.69 and
f(${\phantom{}}^{65}$Cu)=0.31.  

{\it Neutron-capture elements:}
We derived Sr, Y, Zr, Ba, La, Ce, Pr, Nd, Sm, Eu and Dy abundances from the MIKE
spectra, and Y, Ba, La, Ce, Pr, Nd, Sm, Eu from the GRACES spectra.
For all these elements we performed a spectral synthesis analysis, as
hyperfine and/or isotopic splitting and/or blending features needed to
be taken into account. 
In all the cases we have assumed the Solar-system isotopic fractions.

The inferred chemical abundances are listed in
Tabs.\ref{tab:abu1}--\ref{tab:abu2}.  
Internal uncertainties in chemical abundances due to the adopted model
atmospheres were estimated by varying the stellar parameters, one at a
time, by the amounts derived in Section~\ref{sec:atm}, namely
\teff/\logg/[Fe/H]/\vmicro=$\pm$50\,K/$\pm$0.15\,cgs/$\pm$0.10\,dex/$\pm$0.20\,\kmsec.  
Another way to evaluate how the adopted model atmospheres affect
our results is by comparing chemical abundances for the same
stars inferred from the spectroscopic and photometric parameters.
This test is particularly enlightening since we have adopted
spectroscopic parameters for MIKE data and photometric parameters for GRACES.
In the lower panel of Fig.~\ref{fig:box_graces_mike} we plot the
differences $\Delta$[X/Fe] between the chemical abundances obtained
from the photometric and the spectroscopic parameters for the stars n2 and a2
(observed with both MIKE and GRACES) for the elements available from both
the data sets. We note that the differences are generally small, of
the order of a few hundredth dex, and, in any case, lower than 0.10~dex\footnote{Iron is not included in this plot because we treat it separately in Section~\ref{sec:iron}}.
The upper panel represents the differences between the chemical
abundances obtained from GRACES (photometric parameters) and MIKE
(spectroscopic parameters) spectra for the stars n2 and a2. Here the
differences are larger, up to $\sim$0.20~dex.
The arrows displays the amount and direction of the variation in each
species due to the different sets of atmospheric parameters.
We note that, even if in a few cases, e.g.\ Si, the set of
atmospheric parameters can explain the differences in the inferred
abundances, in most cases the distinct results have simply to be ascribed to
the different data sets and to the different spectral lines used in
the analysis. 

In addition to the contribution introduced by internal errors in
atmospheric parameters, we estimated the contribution due to the
limits of our spectra, e.g.\ due to the finite S/N that
affects the measurements of EWs and the spectral synthesis. 
For MIKE spectra, the uncertainty in the EWs measurements has been
estimated to be $\pm$4.4~m\AA\ by comparing the observed distribution
of the differences 
between the Fe lines' EWs for the two MIKE stars and the
corresponding distribution expected from simulated spectra (see
Section~\ref{sec:iron}).  
For GRACES spectra we have compared the EWs obtained from individual
exposures of the same stars and obtained typical errors in EWs of
5~m\AA\ and 7~m\AA\ for the two {\it normal} and the two {\it
  anomalous} stars (the latter having lower S/N), respectively.
For each spectrum, the errors in chemical abundances due to the EWs
have been calculated by 
varying the EWs of spectral lines by the corresponding uncertainty.
For the species inferred from spectral synthesis we have varied the
continuum at the $\pm 1~\sigma$ level, and re-derived the chemical
abundances from each line. 
Since the EWs/continuum placement errors are random, the error 
associated to those elements with a larger number of lines is lower.
Hence, the corresponding uncertainty associated with Fe~\,{\sc i} is
negligible (0.01~dex), while for those species inferred from one or
two spectral lines, the error due to the limited S/N is dominant
(e.g.\, Al, or Ce and Pr for GRACES spectra).
All the error estimates are listed in Tab.~\ref{tab:err} for both the MIKE
spectra and the GRACES spectra, separately in the latter case for the
normal and anomalous stars.

\section{The chemical composition of  NGC\,6934}\label{sec:abb}

The mean metallicity obtained from our stars in NGC\,6934 is
[Fe/H]=$-$1.44$\pm$0.13~dex (rms=0.13~dex) from the two MIKE stars
(spectroscopic parameters),
and [Fe/H]=$-$1.37$\pm$0.12~dex (rms=0.12~dex) from the four stars
observed with GRACES (photometric parameters).
Figure~\ref{fig:abundances} shows a summary of the other chemical
abundance ratios 
obtained for the two stars analyzed with MIKE (upper panel), and for the
four stars analyzed with GRACES (lower panels).

As is typical for Population~II, the NGC\,6934 stars are $\alpha$-enhanced.
More specifically, the chemical abundances relative to Fe for the
stars on the {\it normal} and {\it anomalous} RGB are the same 
within the observational errors for all element groups, i.e.\ $\alpha$,
light, Fe-peak, and $n$-capture elements.
In the next few subsections we consider and discuss all the
interesting abundance patterns we observe in NGC\,6934, focusing on
those elements which play an important role in the multiple stellar
populations phenomenon. We start with iron, and then discuss the
$p$-capture elements and finally the $n$-process elements. 

\subsection{Iron abundances}\label{sec:iron}

Even though our sample of stars is small, the results immediately
suggest the presence of Fe internal variations in NGC\,6934.
From a simple comparison between the spectra of the {\it normal} and
{\it anomalous} stars n2 and a2, which have very similar atmospheric
parameters, it is clear that a2 displays stronger lines, as shown in
Fig.~\ref{fig:Hyd}. On the other hand, the hydrogen lines (H$\alpha$
and H$\beta$) are very similar, confirming that the stars have the
same effective temperature.

Figure~\ref{fig:fe} displays the Fe~{\sc i} and Fe~{\sc ii} abundances
as a function of \x\ for the analyzed stars, for both the MIKE and GRACES
spectra. 
We note the systematic higher Fe abundances for the
GRACES data, as discussed in Section~\ref{sec:atm}, but
to avoid these systematic differences in our discussion,
we will refer to the relative Fe differences observed in each data
set, separately.

In particular, the iron abundances, both neutral and singly ionized, are
higher in the stars located on the {\it anomalous} RGB. 
This remains valid independent of which set of atmospheric  
parameters are used, spectroscopic or photometric.
These {\it anomalous} stars also 
have higher \x\ values than {\it normal} RGB stars in the chromosome map.

In all the panels, two error bars have been associated with each Fe
measurement: the error defined as $\rm {r.m.s./\sqrt{(N-1)}}$, where
$N$ is the number of lines (gray error bars); and the expected error
due to the model 
atmospheres and the S/N (black error bars, as discussed in
Section~\ref{sec:abundances}). 
The larger number of available Fe~{\sc i} lines (upper panels) translates in a
statistical error $\rm {r.m.s./\sqrt{(N-1)}}$ smaller than the one
associated with Fe~{\sc ii} (lower panels). 
The expected errors listed in Tab.~\ref{tab:atm} are larger, ranging between
0.09 to 0.11 for Fe~{\sc i}, and between 0.12 to 0.18 for Fe~{\sc ii}. 

Quantitatively, we obtain $\Delta$[Fe~{\sc i}/H]=0.18$\pm$0.13~dex for MIKE
(by construction $\Delta$[Fe~{\sc ii}/H] is the same, but has a
larger error of $\pm$0.16); and
$\Delta$[Fe~{\sc i}/H]=0.21$\pm$0.15~dex, $\Delta$[Fe~{\sc ii}/H]=0.23
$\pm$0.18~dex, for GRACES.
The weighted averages of the two {\it normal} and {\it anomalous} stars
observed with GRACES are 
$<$[Fe~{\sc i}/H]$>$=$-$1.47$\pm$0.08~dex and
$<$[Fe~{\sc i}/H]$>$=$-$1.26$\pm$0.07~dex, respectively which means
$\Delta$[Fe~{\sc i}/H]=0.21$\pm$0.11~dex. 
By taking advantage of the measurements from both MIKE and GRACES, the weighted
$\Delta$[Fe~{\sc i}/H] mean value is $\Delta$[Fe~{\sc
  i}/H]=0.20$\pm$0.08~dex, which is a significance of $\sim$2.5~$\sigma$.
  
Although the small observed sample prevents us from drawing strong conclusions
on the higher Fe abundance for the {\it anomalous} RGB stars, the Fe
enrichment is observed in all our available measurements,
independent of the set of the adopted stellar parameters and/or data. Thus, it
is very tempting to conclude that the {\it anomalous} stars are enriched
in the overall metallicity. 
In the following we present a few independent tests to corroborate the
presence of Fe variations in NGC\,6934.

First, in Fig.~\ref{fig:ews_fe} we show the difference in EWs for the Fe
lines between the two stars observed with MIKE (a2$-$n2). 
The observed differences have been compared with those expected for
two simulated spectra with the same atmospheric parameters of the two
target stars. We note that, while the observed mean difference for Fe~{\sc i}
is relatively high, 9.9~m\AA, that for the Fe~{\sc ii} is smaller,
2.9~m\AA. In both cases the observed differences agree well with
those expected from theoretical spectra ($\Delta$EWs(Fe~{\sc
  i})=10.4~m\AA\ and $\Delta$EWs(Fe~{\sc ii})=1.2~m\AA). 
The small difference in the Fe~{\sc ii} lines is well explained by
the atmospheric parameters of our stars. Indeed, we get higher surface
gravities for the two stars located on the chromosome map's
{\it anomalous} sequence. And, assuming that all the other atmospheric
parameters are identical, a star with higher \logg\ is expected to
show smaller EWs.

Second, an independent test that provides further evidence for iron variations
in NGC\,6934 comes from the comparison between the observed CMD and
stellar isochrones. 
Figure~\ref{fig:iso}
represents the $m_{\rm F336W}$-$(m_{\rm F336W}-m_{\rm F814W})$ and
$m_{\rm F438W}$-$(m_{\rm F438W}-m_{\rm F606W})$ CMDs for the
cluster. The presence of an additional sequence, on the red side of
the main RGB, is clear in the $m_{\rm F336W}$-$(m_{\rm F336W}-m_{\rm
  F814W})$ CMD\@. Stars in the {\it anomalous} RGB, which {\it
  evolve} from a fainter SGB, correspond to the stars defining the
{\it anomalous} sequence on the chromosome map (on
Fig.~\ref{fig:cmdtarget}), and are also located redder in the $m_{\rm
  F438W}$-$(m_{\rm F438W}-m_{\rm F606W})$ CMD\@.
Superimposed on both
CMDs are isochrones corresponding to an age=12.25~Gyr, Y=0.2471,
[$\alpha$/Fe]=$+$0.40 and [Fe/H]=$-$1.40 (dark red) and [Fe/H]=$-$1.60
(black), retrieved from the Dartmouth database (Dotter et al.\,2008).
Although the upper part of the RGB is poorly represented by these
isochrones, especially in the $m_{\rm F336W}$-$(m_{\rm F336W}-m_{\rm
  F814W})$ CMD (possibly due to the contribution of CNO abundances to
$m_{\rm F336W}$), the fit on the lower RGB and the SGB is
satisfactory, with only some systematic shift of both isochrones on the
RGB blue side.  
We conclude that a difference in [Fe/H] by $\sim$0.2~dex can reproduce
the CMD of NGC\,6934. According to our {\it HST} high-precision CMD,
NGC\,6934 hosts one main stellar population, and one minor component
that is slightly enriched in Fe,
and which constitutes 7$\pm$1\% of the cluster mass (Milone et al.\,2017).

The inset in Fig.~\ref{fig:iso}, shows the theoretical \logg-\teff\
plane from the Dartmouth isochrones with [Fe/H]=$-$1.60 and
[Fe/H]=$-$1.40. Superimposed are our adopted atmospheric parameters
for the analysed stars, specifically spectroscopic parameters for MIKE
and photometric parameters for GRACES (see Section~\ref{sec:atm}). Both
sets of adopted \logg-\teff\ parameters agree with the parameters
expected from theoretical isochrones and satisfy the basic principle
that at a given \teff\, stars with higher metallicity have higher \logg.
Specifically, the \logg\ difference expected at \teff$\sim$4450~K is
$\sim$0.15-0.20~dex, consistent with our estimates.

\subsection{Light elements}

The abundance ratios obtained for the $p$-capture elements from C to Al have
been plotted in the left panel of Fig.~\ref{fig:light-s}. As a
comparison, we have 
also included the same elements derived for M\,22 by Marino et
al.\,(2009, 2011). 
A first fact to note is that the analyzed stars in NGC\,6934 do
not seem to show large dispersions in these elements. 
All our NGC\,6934 targets lie in the O-rich/Na-Al-poor range,
e.g.\ they all share a primordial chemical composition as regards
the $p$-capture elements. On the other hand, both the $s$-poor and the
$s$-rich stars analysed in M\,22 span a relatively large range of
abundances in all these elements, including stars with chemical
composition typical for 
second-generation(s), e.g.\ O-depleted and Na-Al-enhanced.
Carbon abundances have been inferred only for the two stars observed
with MIKE, and the {\it normal} and the {\it anomalous} stars show
similar values.

The lack of second-generation(s) stars in our NGC\,6934  sample is due to the
selection of our targets, which is biased towards stars located in the
lower part of the chromosome map with relatively low \y\ values. As
shown in Milone et al.\,(2017), first population stars possess relatively
low values of \y, and relatively high values of \x. 
The choice to
observe only stars in the lower part of the chromosome map is
justified by our goal to investigate 
any difference (if present) between the chemical composition of stars
in the {\it normal} and {\it anomalous} RGBs. To do this, we tried to avoid 
additional effects such as the variation in light elements
within the {\it normal} and {\it anomalous} categories. 
We expect that second population(s) stars, with low O and high Na,
would be observed to have higher \y\ values. 
In this context, the higher Na and Al abundance ratios inferred for the {\it
normal} star n1 agree with its position on the chromosome map, given its
slightly higher \y.
Future observations will enlighten the light elements distribution in
the {\it normal} and {\it anomalous} RGBs of NGC\,6934.

\subsection{$s$-process elements}

On the right panel of Fig.~\ref{fig:light-s} we show the abundance
ratios of some 
of the analyzed $n$-capture elements, specifically Y, Ba, La and Nd
relative to Fe, as a function of [Fe/H] both from MIKE and GRACES
spectra. As done for the light elements, in each panel we plot the
corresponding results obtained for M\,22 (data from Marino et
al.\,2009; 2011). 

Although it is a more metal poor GC, M\,22 has been chosen for comparison
purposes because it shows a clear bimodality in $n$-capture
elements. 
Typically, in {\it anomalous} GCs, like M\,22, the variation in
$n$-capture elements is due to additional $s$-process enrichment among some
second generation stars (Marino et al.\,2011). 
Specifically, M\,22 hosts a stellar population relatively enriched in
the $s$-process elements 
($s$-rich group) with respect to a stellar population with lower
$s$-elements abundance ratios ($s$-poor group). The $s$-rich group is
also enhanced in Fe by $\sim$0.15~dex and in
the overall [C+N+O/Fe] (Marino et al.\,2009, 2011, 2012).  However,
the $n$-capture 
element [Eu/Fe] ratio, which is mostly produced via {\it 
  rapid} neutron capture reactions ($r$-process element), is uniform.  

We note that the difference in Fe between the $s$-rich and the
$s$-poor groups of M\,22 is similar to the measured difference between
the {\it normal} and {\it anomalous} stars in NGC\,6934.
However, while for M\,22 the variation in Fe is coupled with a variation in
[$s$-elements/Fe], with the Fe-richer stars displaying higher
[$s$-elements/Fe], the analyzed stars in NGC\,6934 exhibit similar
content for Y, Ba, La, and Nd relative to Fe. 
Indeed, by considering, e.g.\ La, which is one of the $s$-elements with
more precise measured abundances, we derive a difference of
$|~\Delta$[La/Fe]~$|$=0.05$\pm$0.11~dex (from MIKE data)
and $|~\Delta$[La/Fe]~$|$=$-$0.10$\pm$0.11~dex (from GRACES data), 
compared to the
significantly larger difference
$|~\Delta$[La/Fe]~$|$=0.32$\pm$0.02~dex between the $s$-rich and $s$-poor
group of M\,22.
We recall here that M\,22 displays one of the smallest internal variation in the
[$s$-element/Fe] ratio among the known {\it anomalous} GCs (see e.g. Fig.~19 in
Marino et al.\,2015). For comparison, M\,2
has $|~\Delta$[La/Fe]~$|$=0.58~dex (Yong et al.\,2014), NGC\,5286
$|~\Delta$[La/Fe]~$|$=0.56~dex (Marino et al.\,2015), and
NGC\,6273 $|~\Delta$[La/Fe]~$|$=0.42~dex (Johnson et al.\,2017).

Clearly, the Type~II GC NGC\,6934 seems 
different from the
chemically-defined class of {\it anomalous} GCs, as our analysed
sample does not show any evidence of additional internal variations in
$s$-elements.  
This conclusion is based only on
  four stars, and future observations on larger sample sizes may
  perhaps reveal the presence of stars enhanced in $s$-elements located in other
regions of the chromosome map.
On the other hand, we note that, despite our sample of only four
stars, they have been carefully selected from the two different
sequences in the chromosome map where we expect variations in heavy
elements, as observed in the {\it anomalous} GCs. 
Photometrically, $s$-rich stars in {\it anomalous} GCs are located on
the red RGB (Milone et al.\,2017).  
We cannot exclude, however, a different behavior for NGC\,6934.

An inspection of the other $n$-capture species in
Fig.~\ref{fig:abundances} (from Sr to Dy for MIKE
and from Y to Eu for GRACES) clearly suggests the lack of any significant
internal variation in the analysed elements in NGC\,6934.
In this figure we note a growth of the abundances relative to Fe as a
function of the atomic number in both the {\it normal} and {\it
  anomalous} stars. 

\section{Discussion and conclusions}\label{sec:conclusions}

We have presented a high-resolution chemical abundance analysis of
four stars on the {\it normal} and {\it anomalous} RGB of NGC6934.
We have found that the chemical abundances of all the inferred
species, except Fe, are consistent with uniform abundance ratios
in the four 
analyzed stars. The difference in Fe is of the order of $\sim$0.2~dex
with two {\it anomalous} RGB stars exhibiting higher abundances. 
Such variation in Fe is corroborated by the comparison between the
CMDs obtained from high-precision $HST$ photometry and the isochrones
of Dotter et al.\,(2008). 

NGC~6934 has been classified as a Type~II GC as, contrary to Type~I GCs, 
it displays more than the normal 1G and 2G sequences in the
\x-\y\ photometric plane. 
Milone et al.\,(2017) have shown that the {\it main} pattern on the
chromosome map in all GCs, both Type~I and Type ~II, corresponds
to a distribution in light element abundances: stars with different
He, C, N, O, and Na abundances
distribute from higher \x\ and lower \y\ to lower \x\ and higher \y. 

The presence of numerous additional sequences on the red side of the main
chromosome map appears in $\sim$18\% of the analyzed GCs (Type~II).
For a subsample of these GCs, it was possible to identify some stars
with available chemical abundances along the sequences of the
chromosome map, and it appeared that the additional red sequences
are populated by stars enriched in both Fe and $s$-elements relative
to Fe ({\it anomalous} GCs). This behavior is observed also in
$\omega$~Centauri. 

For simplicity, a list of the currently known chemically {\it anomalous}
and/or photometrically peculiar Type~II GCs is provided in Tab.~7. 
In the light of the new findings coming from the synergy between
spectroscopic and photometric investigation, we propose a slighly
different and simpler terminology to define this new class of GCs, as
listed in the last column of the table:
\begin{itemize}
\item{Iron~II are the objects with variations in Fe, in contrast to
    the typical GCs, designated Iron~I, that do not show evidence for
    star-to-star 
    variations in Fe;}
\item{$s$~II is used for the GCs with variations in $s$-elements,
    while typical GCs are labeled $s$-I;}
\item{following Milone et al.\,(2017), Type~II is used for the GCs
    with additional red sequences on the chromosome map, while typical
  GCs are Type~I.}
\end{itemize}
By using this terminology, NGC\,6934 can be classified as a
Iron~II/$s$-I/Type~II GC. 
However, we cannot exclude that a future analysis of larger sample
sizes may reveal stars with enhancements in the $s$-elements. 

Our results show that, although NGC\,6934 is classified as a Type~II
GC, its {\it anomalous} stars do not exhibit any enrichment in the
$s$-process elements relative to Fe, although variations in the [Fe/H]
abundances are present.
At present, NGC\,6934 is the only Type~II GC with available
spectroscopy on the {\it red} sequence of the chromosome map, that displays
chemical variations in Fe, but not in the $s$-elements\footnote{Yong et
  al.\,(2014) has shown that the extremely Fe-rich population in M\,2,
  contrary to the Fe-intermediate one,
  does not show any enhancement in the [$s$-elements/Fe]. However the
extremely Fe-rich stars have other chemical peculiarities, such as
lower [Ca/Fe] abundances.}.  
A larger sample of stars would help to assess if the Fe distribution
in NGC\,6934 is continuous or consistent with two discrete populations.

Another photometric similarity between NGC\,6934 and {\it anomalous}
GCs with chemical variations in [$s$-elements/Fe] is the split SGB, as
shown in Figs.~\ref{fig:cmdtarget} and \ref{fig:iso}. In the {\it
  anomalous} GCs this feature has been associated with the presence
of stellar populations with different C+N+O (Marino et al.\,2011;
2012). 
As an example, the comparison between isochrones and the CMD of M22,
reveals that a variation in Fe alone could not reproduce the observed
SGB split (Marino et al.\,2009). 
In contrast, isochrones with the same age and with a metallicity
difference of 0.2~dex, corresponding 
to the observed Fe difference between the {\it normal} and {\it anomalous} 
stars, can account for the size of the
SGB split in NGC\,6934, without invoking any variation in the overall C+N+O. 
Although we cannot estimate the total C+N+O for NGC\,6934 without N
abundance measurement,
the lack of additional internal variations in $s$-process elements points to no
strong intra-cluster pollution by low-mass asymptotic giant branch
stars that underwent third dredge-up (at least not at the level
observed in the $s$~II GCs); and might in turn suggests no 
variations in the [C+N+O/Fe] (e.g.\, Karakas\,2010). 

We emphasize here that the shape of the distribution of the {\it
  anomalous-red} stars in 
the chromosome map of NGC\,6934 shows a few differences
from that observed in other {\it anomalous} GCs. While, most of the
{\it red} stars in 
the {\it anomalous} GCs, such as M\,22, have relatively high
values of \y, suggesting an enrichment in light elements, in NGC\,6934
they are located on the lower part of the diagram (see
Fig.~\ref{fig:n6934_m22_n6752}). 
Even though we think it is very unlikely to have
$s$-enriched stars at higher \y\ with a lack of them at lower values, we
cannot exclude it {\it a priori}. 
Future observations sampling a larger number of stars on all
the stellar populations appearing in the chromosome map of NGC~6934 
will allow us to better constrain the extent of the Fe variations in this GC.

Our chemical abundance analysis suggests that there is no one-to-one
correlation between the appearance of {\it red}
additional sequences on the chromosome maps of Type~II GCs and
variations in $s$-elements and, possibly C+N+O. Could these
sequences, instead, be more directly linked to enrichments in the overall
metallicity? This idea opens a new perspective in the interpretation of
the origin of $\sim$18\% of the Milky Way GCs. Could all these objects
have been significantly more massive than typical GCs in order to
support a more prolonged star formation? Could the retention of SNe 
ejecta hint that these objects originated in a
deeper potential well, such as the nucleus of a dwarf galaxy (Da Costa\,2015)?
Future observations will shed light on the answers to these questions.

\acknowledgments
The authors thank the anonymous referee for useful discussion.

Australian access to the Magellan Telescopes was supported through the
National Collaborative Research Infrastructure Strategy of the
Australian Federal Government. 

Based on observations obtained with ESPaDOnS, located at the
Canada-France-Hawaii Telescope (CFHT). CFHT is operated by the
National Research Council of Canada, the Institut National des
Sciences de l'Univers of the Centre National de la Recherche
Scientique of France, and the University of Hawai'i. ESPaDOnS is a
collaborative project funded by France (CNRS, MENESR, OMP, LATT),
Canada (NSERC), CFHT and ESA. ESPaDOnS was remotely controlled from
the Gemini Observatory, which is operated by the Association of
Universities for Research in Astronomy, Inc., under a cooperative
agreement with the NSF on behalf of the Gemini partnership: the
National Science Foundation (United States), the National Research
Council (Canada), CONICYT (Chile), Ministerio de Ciencia, Tecnolog\'{i}a e
Innovación Productiva (Argentina) and Minist\'{e}rio da Ci\^{e}ncia,
Tecnologia e Inova\c{c}\~{a}o (Brazil). 

AFM, GDC and HJ acknowledge support by the Australian Research Council through
Discovery Early Career Researcher Award DE160100851 and Discovery
projects DP150103294 and DP150100862. 
APM has been supported by the European Research Council through the
Starting Grant ``GALFOR''.

\facilities{Magellan:Clay, Gemini:Gillett, HST}

%
   \begin{figure*}
\center
   \includegraphics[width=18.2cm]{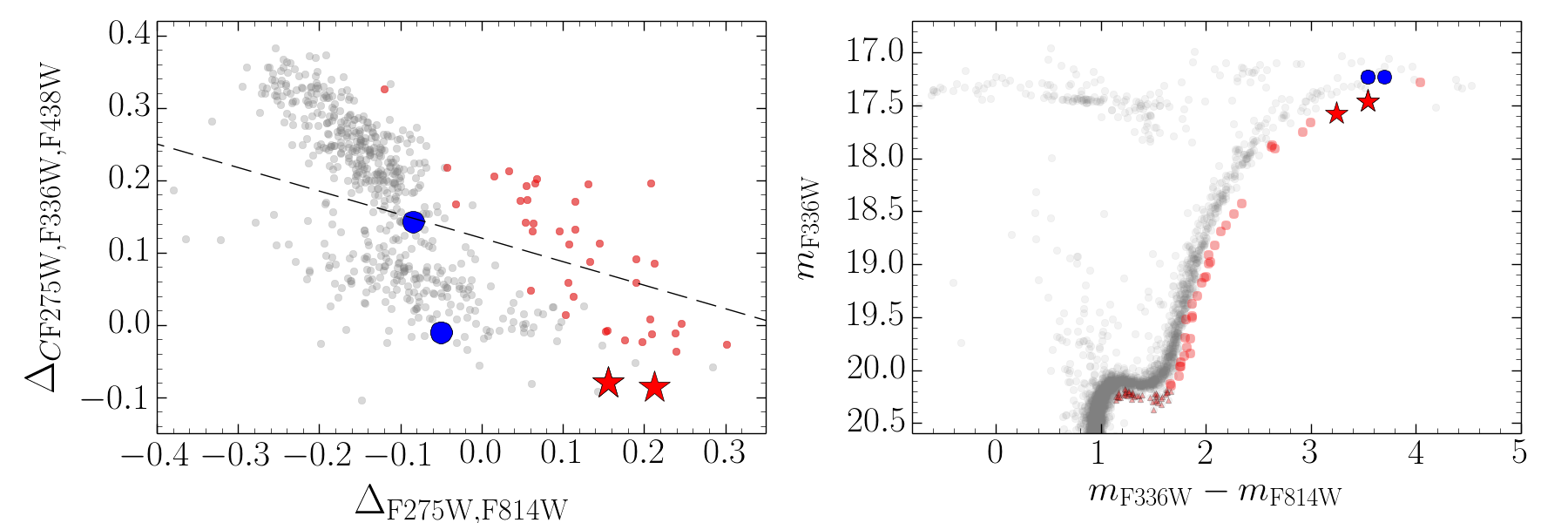}
      \caption{
       Location of the spectroscopic targets in the chromosome map
       (left panel) and the $m_{\rm F336W}$ vs. $(m_{\rm F336W}-m_{\rm
         F814W})$ CMD (right panel) of NGC\,6934. 
       The stars coloured in red on the chromosome map have been
       selected as the stars defining the
       redder RGB stars on the CMD in the right panel (red circles).
       Gray symbols represent {\it normal} RGB
       stars, which are observed in both Type~I and Type~II GCs. 
       The red triangles on the CMD represent the faint SGB stars
       of NGC\,6934.
       The dashed black line on the map is the same as used in Milone et
       al.\,(2017) to separate 2G from 1G stars.
       Spectroscopic targets located on the {\it normal} and {\it anomalous}
       RGBs are shown as blue circles and red star-like symbols, respectively. 
       }
        \label{fig:cmdtarget}
   \end{figure*}
%

%
   \begin{figure*}
   \centering
   \includegraphics[width=17cm]{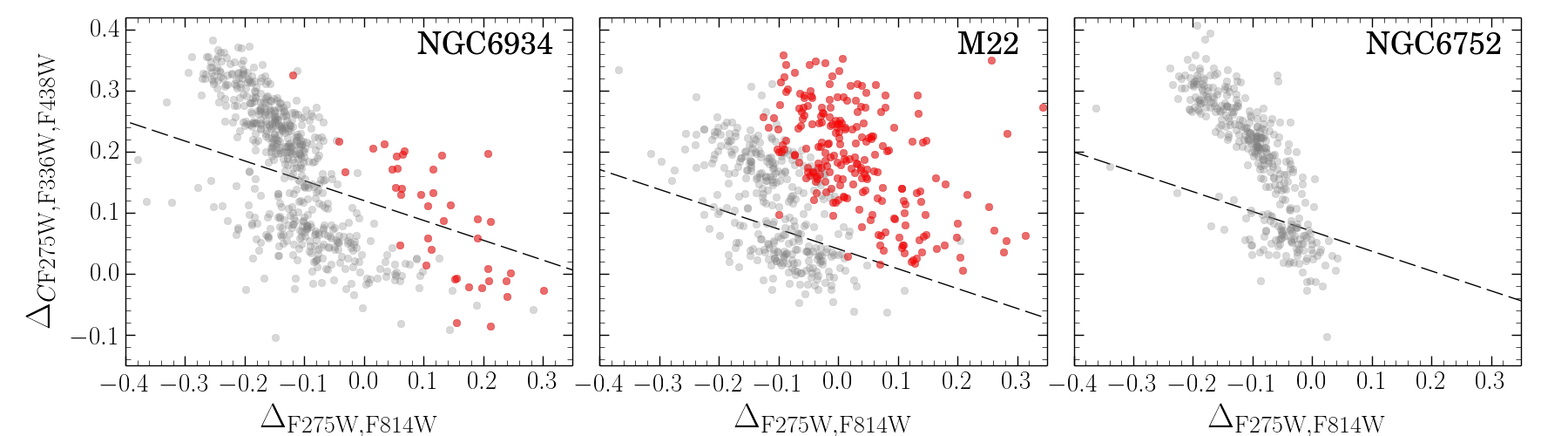}
      \caption{Comparison between the chromosome map of RGB stars in
        the Type~I GC NGC\,6752 (right panel)
        and in the Type~II GCs NGC\,6934 (left) and M\,22 (middle). 
        We colored gray the stars in the {\it normal} RGB, which are
        observed in both Type~I and Type~II GCs. Type\,II GCs exhibit
        an additional sequence of {\it anomalous} RGB stars, which are colored in red.
      In all the panels, the dashed black line separates 2G from
        1G stars, as in Milone et al.\,(2017).
      }
        \label{fig:n6934_m22_n6752}
   \end{figure*}
%
%
   \begin{figure*}
   \centering
   \includegraphics[width=16.5cm]{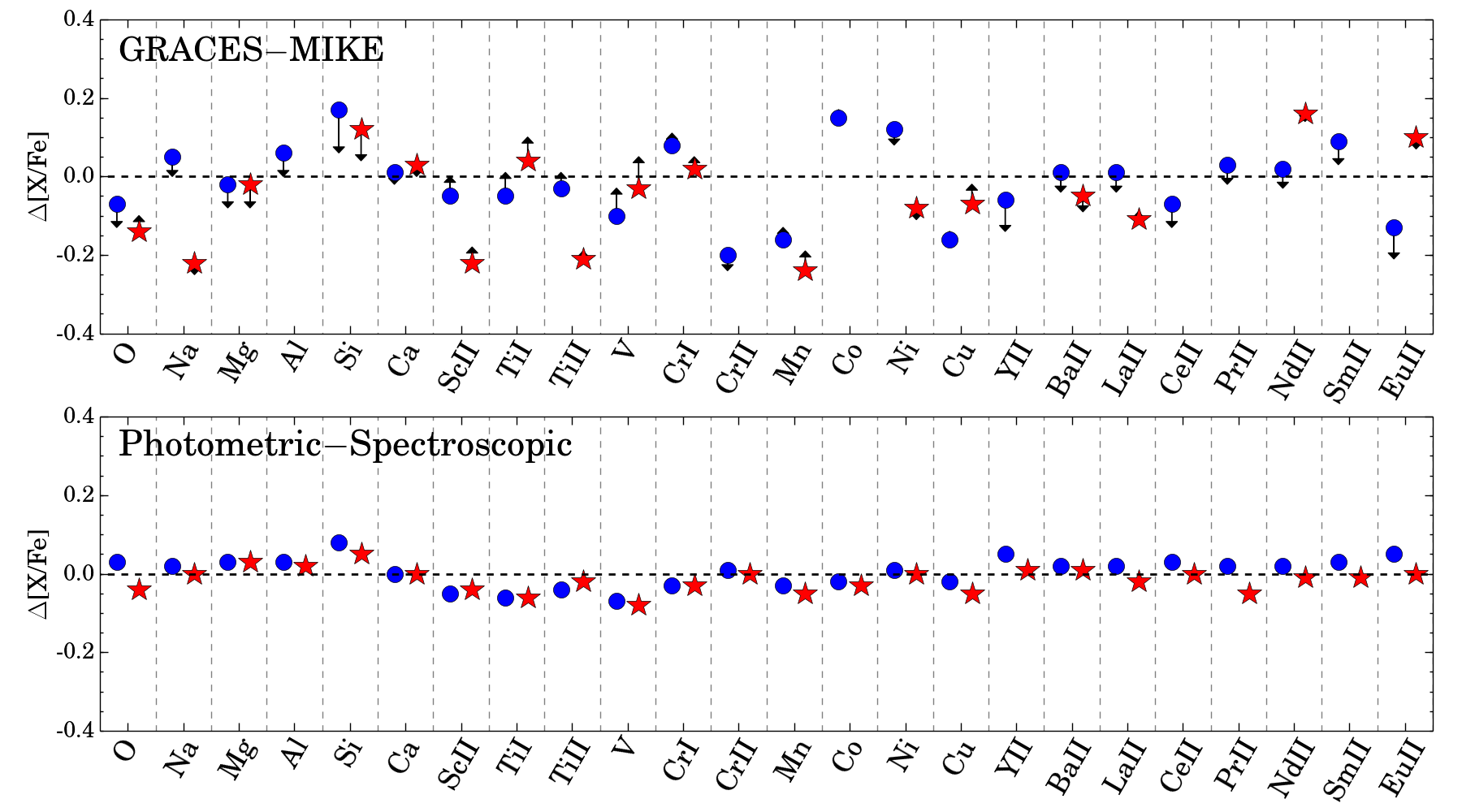}
      \caption{{\it Upper panel}: abundance ratio differences,
        [X/Fe]$_{\rm GRACES} -$[X/Fe]$_{\rm MIKE}$,
        obtained for the stars n2 (blue circle) and a2 (red star),
        observed with both the MIKE and
        GRACES spectrographs.  
        {\it Lower panel}: abundance ratios differences, [X/Fe]$_{\rm
          phot} -$[X/Fe]$_{\rm spectr}$, derived for n2
        and a2, observed with MIKE spectra, and analyzed by using
        stellar parameters from photometry ([X/Fe]$_{\rm phot}$) and
        spectral Fe lines ([X/Fe]$_{\rm spectr}$). 
        The offsets ($\Delta$[X/Fe]=[X/Fe]$_{\rm  phot}
        -$[X/Fe]$_{\rm spectr}$) are shown in the upper panel
        with black arrows to visualize possible differences in the
        abundance ratios due to the different sets of atmospheric parameters
        assumed for MIKE and GRACES spectra.
      }
        \label{fig:box_graces_mike}
   \end{figure*}
%

%
   \begin{figure*}
   \centering
   \includegraphics[width=15cm]{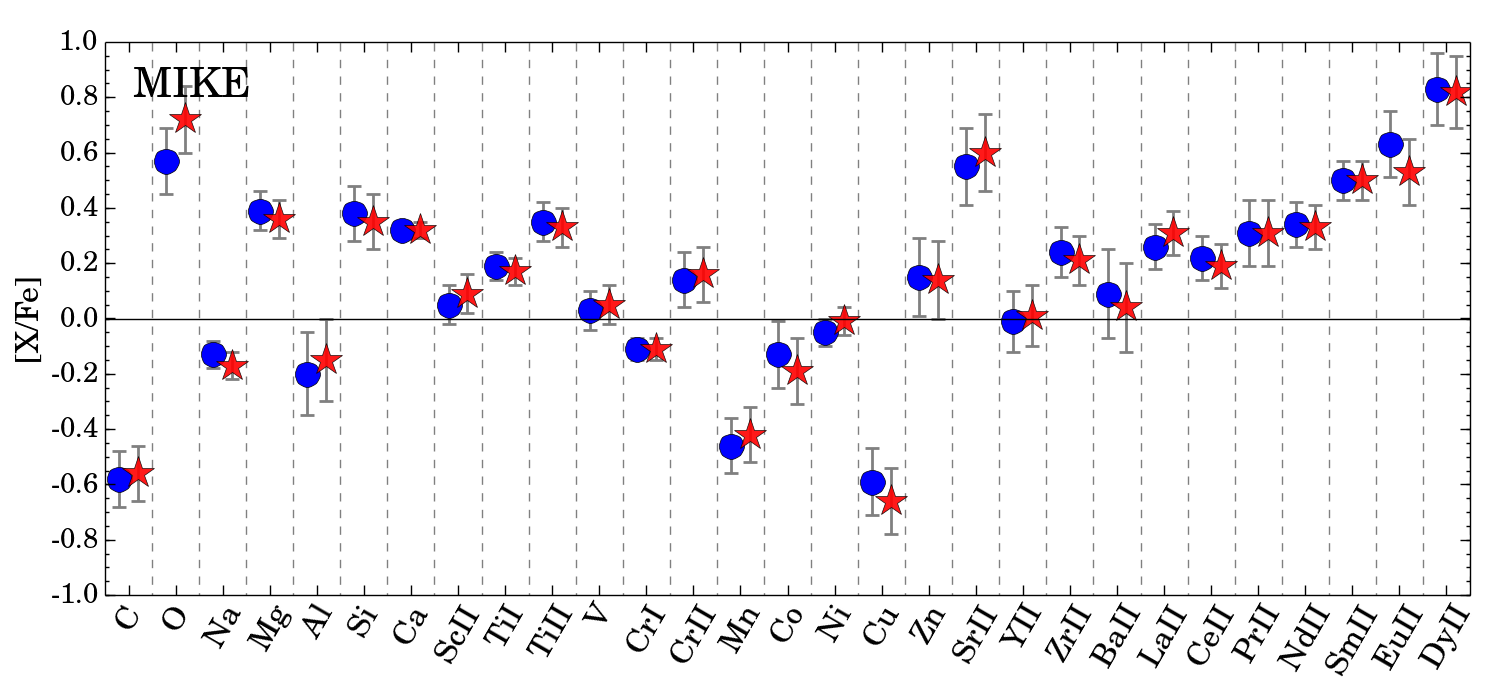}
   \includegraphics[width=15cm]{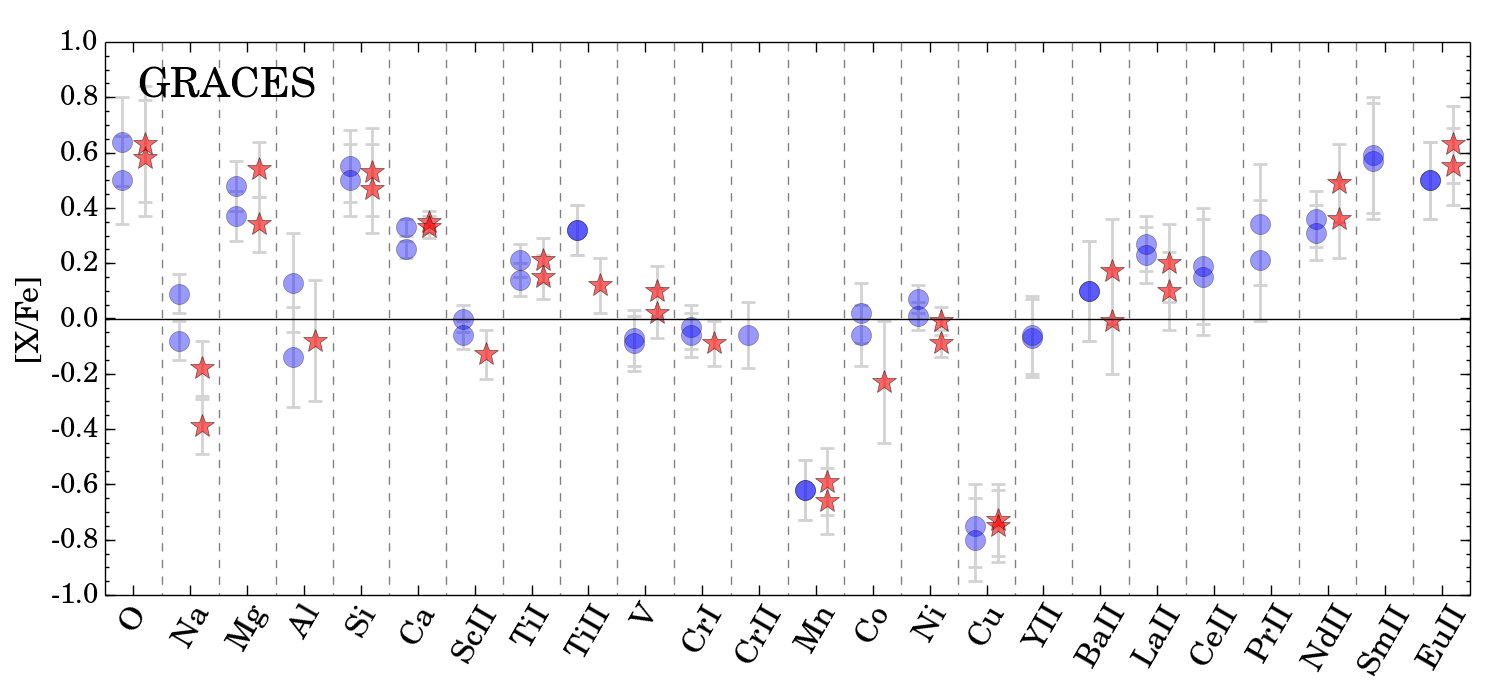}
      \caption{Summary of the abundance ratios results obtained from the
        MIKE (upper panel) and GRACES (lower panel) spectra. For each
        species, we plot the [X/Fe] relative abundances. In both
        panels, red star-like symbols are used for stars located on
        the red sequence of the chromosome map, while, filled blue
        circles are for stars on the {\it normal}
        sequence.} 
        \label{fig:abundances}
   \end{figure*}
%

%
   \begin{figure*}
   \centering
   \includegraphics[width=16cm]{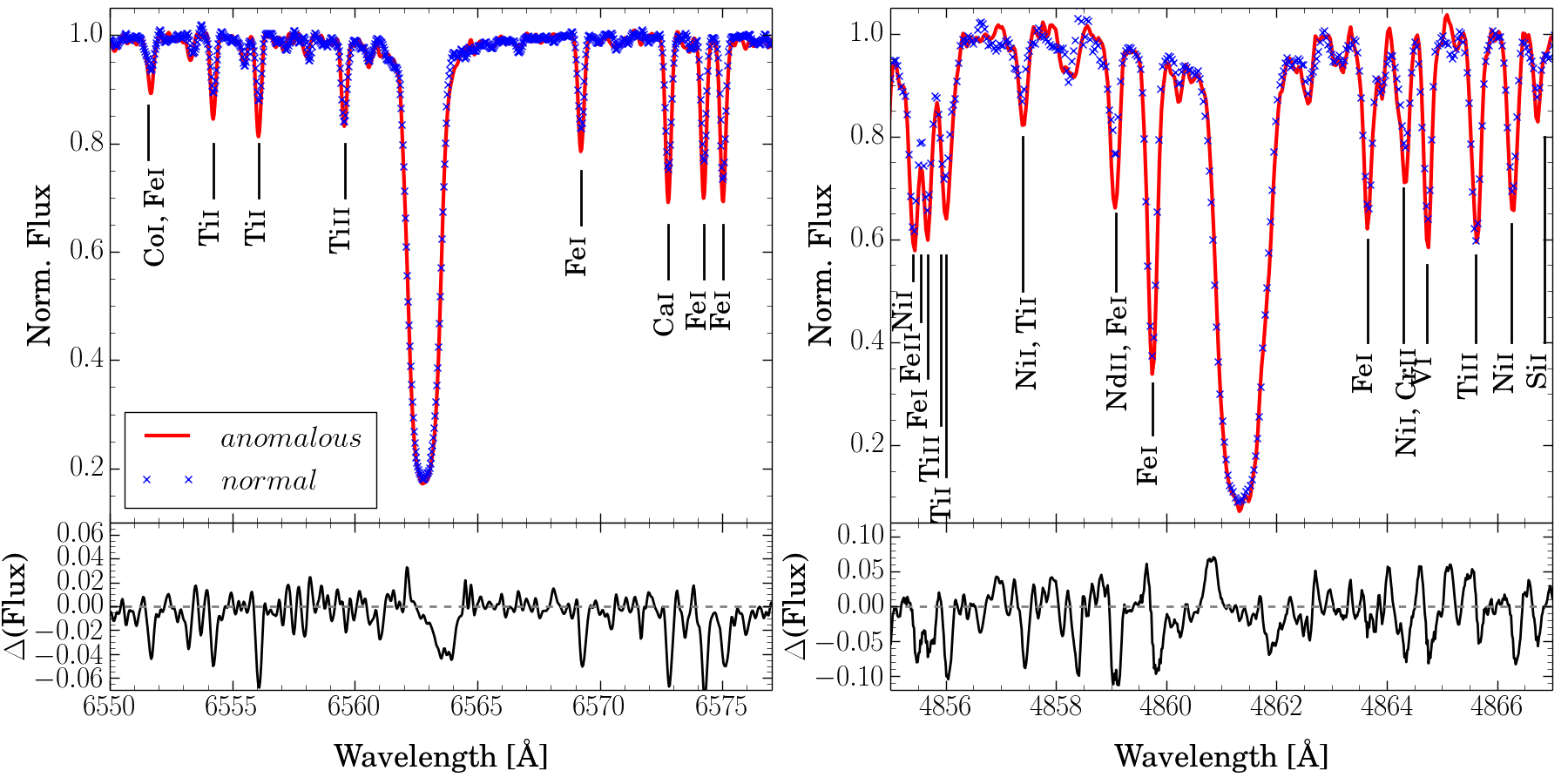}
      \caption{
        Spectral regions around the H$\alpha$ (left) and H$\beta$
        (right) for the two 
        stars observed with MIKE: the star located on the {\it normal}
        RGB is represented with blue
        crosses, while the {\it anomalous} RGB star by a red continuous line. 
        The similarity between the H lines confirms that the two stars
        have very similar atmospheric parameters.
        The difference between the normalized fluxes of the two stars is
        displayed in the lower panels.
       }
        \label{fig:Hyd}
   \end{figure*}
%
%
   \begin{figure}
   \centering
   \includegraphics[width=8.6cm]{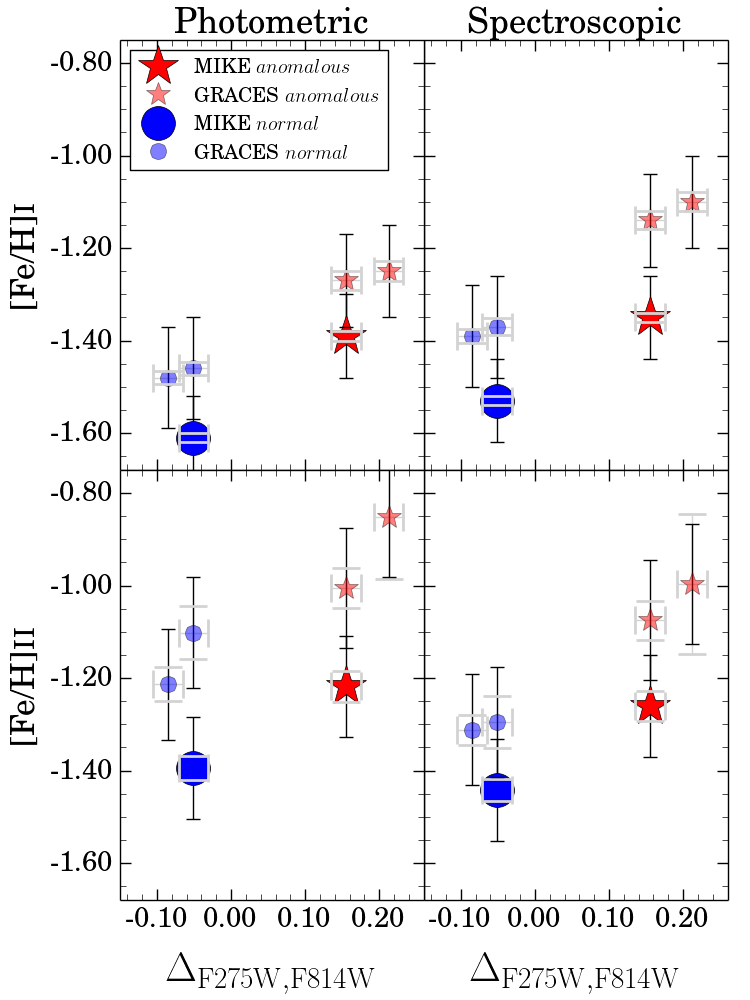}
      \caption{Chemical abundances obtained for Fe\,{\sc i} and
        Fe\,{\sc ii} by using atmospheric parameters derived from Fe spectral
        lines (right panels) and from photometry (left panels). Stars
        in the {\it normal} and red sequence of the chromosome map
        have been plotted with blue circles and red starlike symbols,
        respectively. Results obtained from MIKE and GRACES have been
        distinguished by using larger darker symbols for MIKE, and
        smaller lighter symbols for GRACES.
       }
        \label{fig:fe}
   \end{figure}
%
%
   \begin{figure*}
   \centering
   \includegraphics[width=16.6cm]{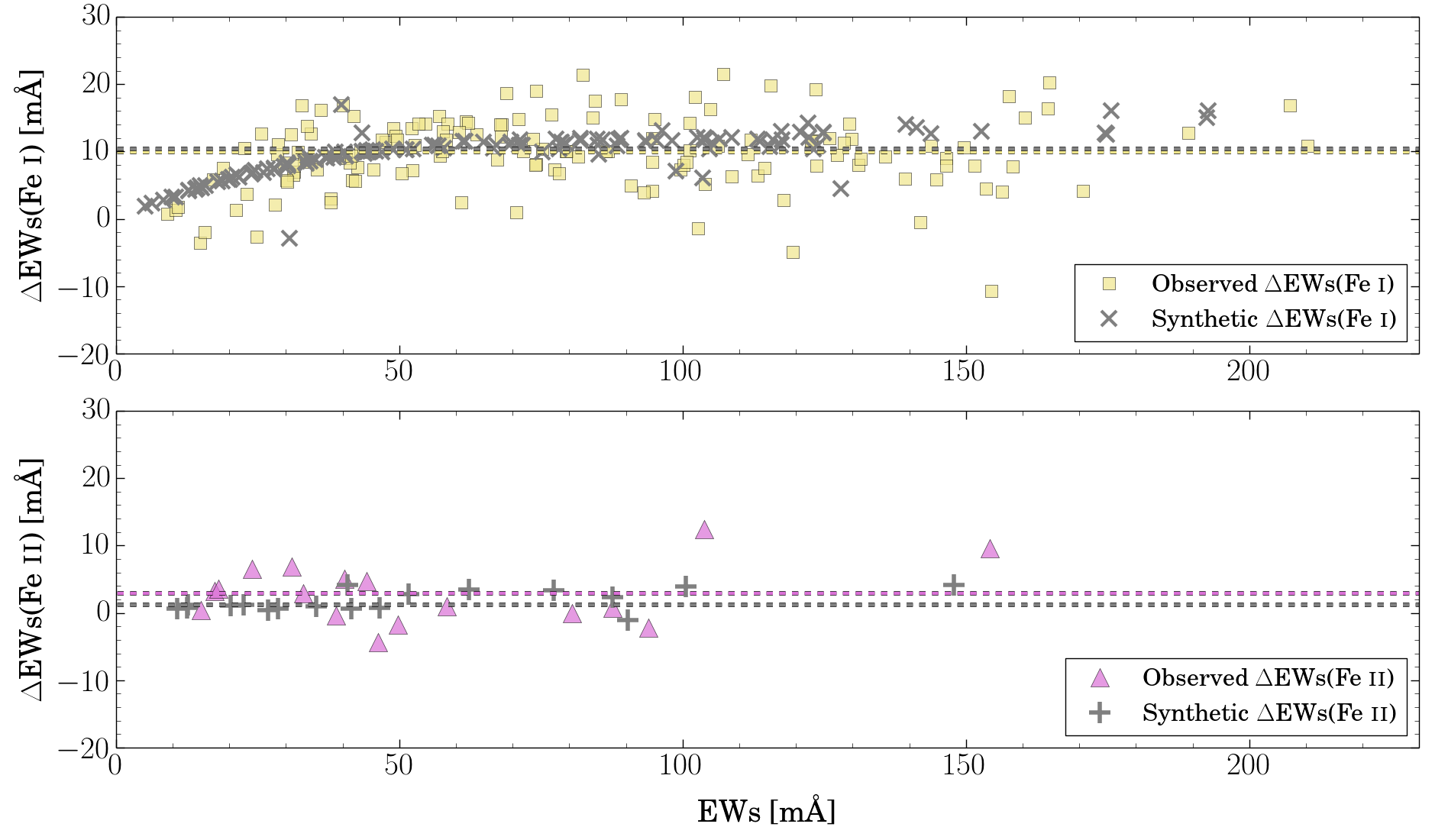}
      \caption{The observed difference in Fe\,{\sc i} (upper panel, yellow
        squares) and Fe\,{\sc ii} (lower panel, magenta triangles) EWs between the stars n2 and a2, observed
        with MIKE, as a function of EW. The two stars have very similar atmospheric
        parameters, but have [Fe/H] differing by $\sim$0.20~dex. Grey symbols
        in both panels represent the EW differences for the same Fe
        lines, but computed from two synthetic spectra with
        the same atmospheric parameters as the two program stars and a
        difference of 0.2 dex in [Fe/H].}
        \label{fig:ews_fe}
   \end{figure*}
%

%
   \begin{figure*}
   \centering
   \includegraphics[width=16.5cm]{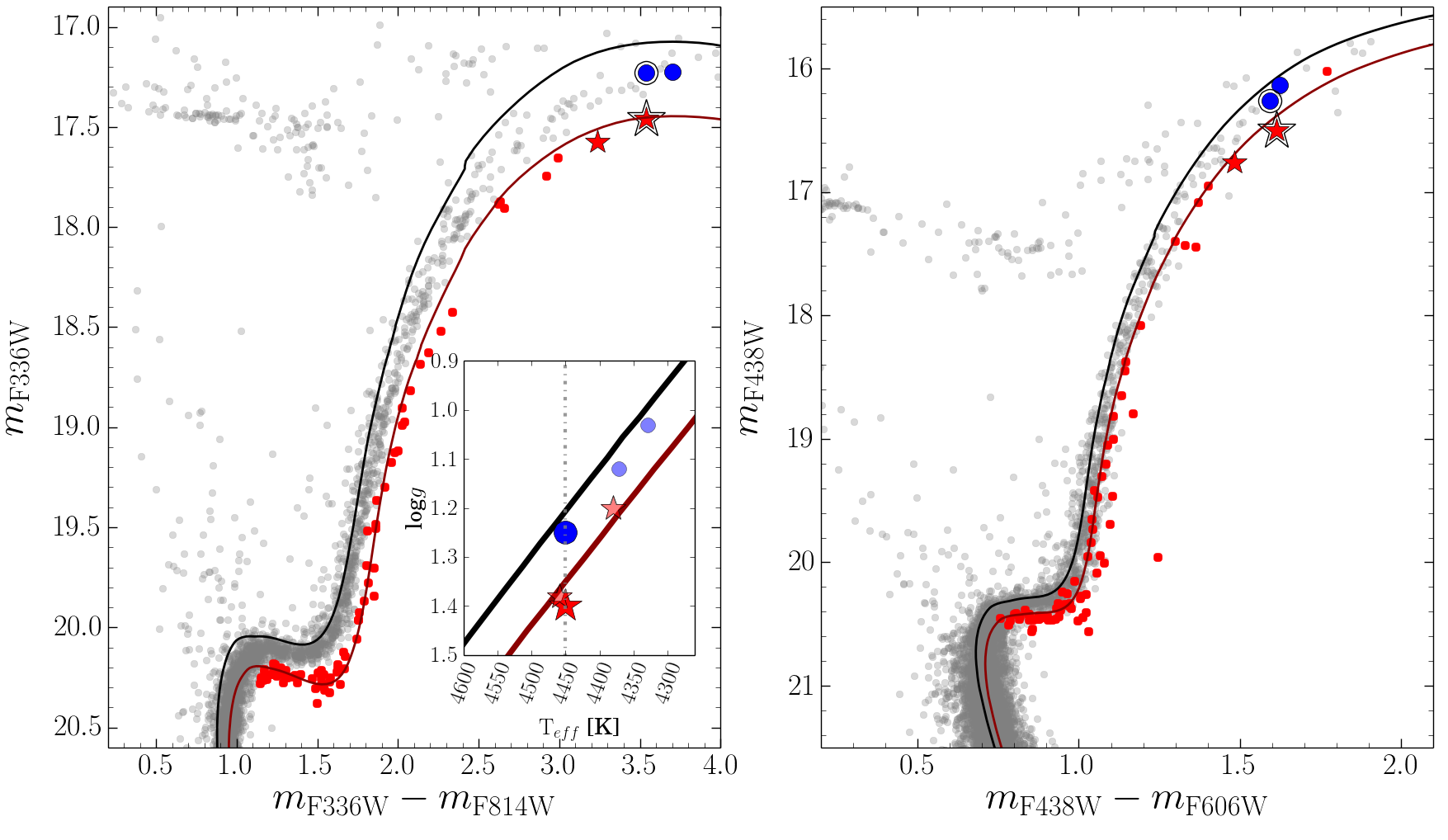}
      \caption{$m_{\mathrm{F336W}}$ versus
        $m_{\mathrm{F336W}}-m_{\mathrm{F814W}}$ (left panel) and $m_{\mathrm{F438W}}$ versus
        $m_{\mathrm{F438W}}-m_{\mathrm{F606W}}$ (right panel) CMDs for NGC\,6934. 
      {\it Anomalous} RGB stars and faint SGB stars have been plotted in red colours.
      Superimposed on the CMDs are two isochrones, from the
      Dartmouth database (Dotter et al.\,2008), with the same age
      (12.25~Gyr), helium (Y=0.2471) and $\alpha$ content 
      ([$\alpha$/Fe]=$+$0.40), but different metallicity. Specifically, the black isochrone has
      [Fe/H]=$-$1.60, and the dark-red one [Fe/H]=$-$1.40~dex. The
      inset on the left panel shows the \logg\, versus \teff\
      theoretical plane with the pattern corresponding to the two
      isochrones, together with the adopted stellar parameters for
      the target stars. The plotted \logg-\teff\ values have been
      derived spectroscopically for the MIKE data
      (plotted with larger symbols), and photometrically for GRACES
      data (plotted with smaller lighter symbols). }
        \label{fig:iso}
   \end{figure*}
%

%
   \begin{figure*}
   \centering
   \includegraphics[width=8cm]{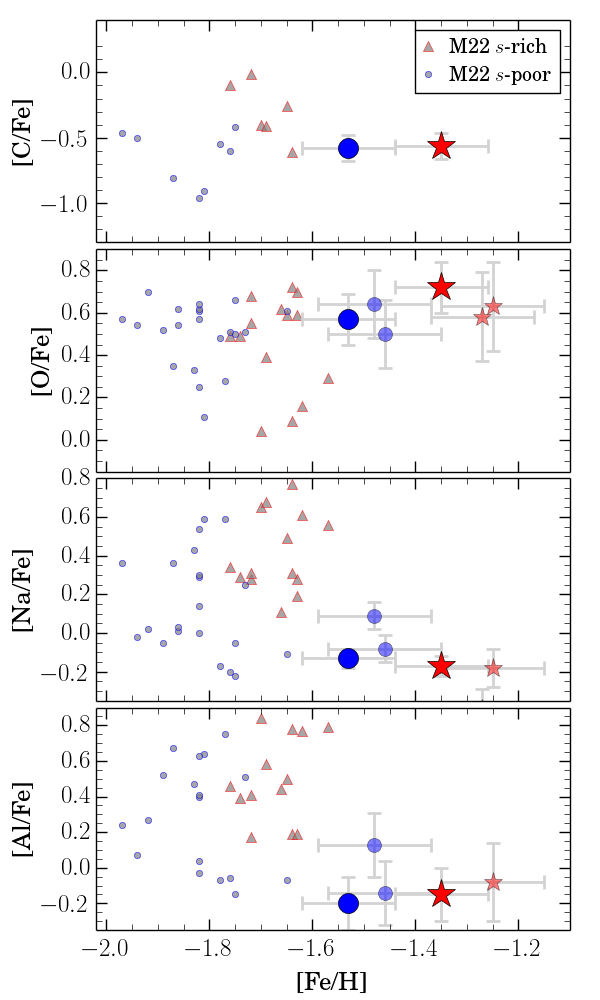}
   \includegraphics[width=8cm]{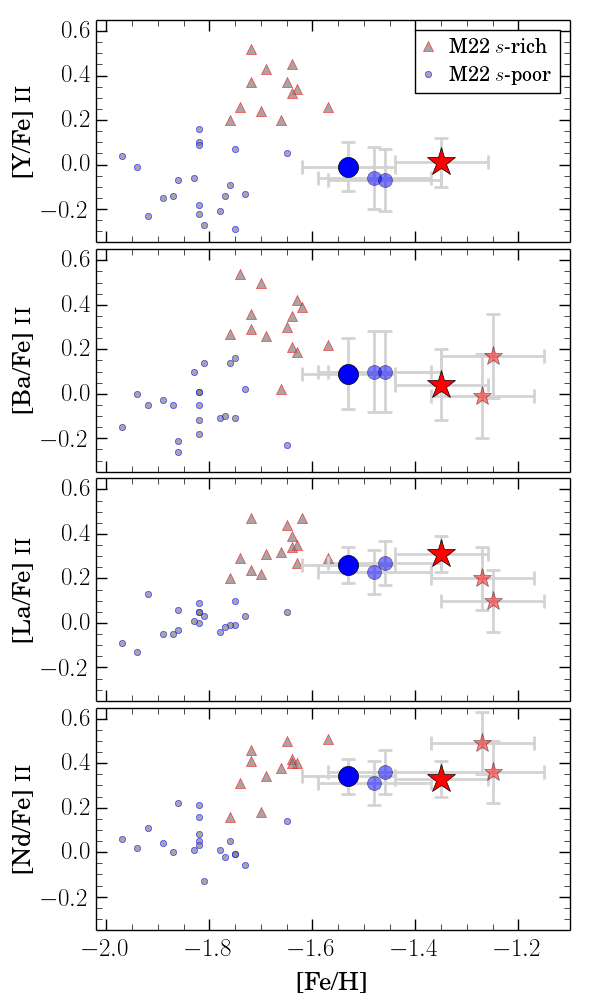}
      \caption{Abundance ratios of light elements [C-O-Na-Al/Fe] (left
        panel) and [Y-Ba-La-Nd/Fe]\,{\sc ii} (right panel),
        derived for NGC\,6934 as function of [Fe/H], compared to the
        two $s$ groups observed 
        in M\,22. $s$-poor and $s$-rich stars in M\,22 are
        plotted with small grey-blue circles and grey-red triangles,
        respectively. Due to the different adopted O solar abundance,
        [O/Fe] abundances for M\,22 have been shifted by $+$0.14~dex.
        Blue circles and red star-like symbols represent
        stars in the {\it normal} and red sequence of the NGC\,6934
        chromosome map, respectively.       }
        \label{fig:light-s}
   \end{figure*}
%

%
%
\begin{deluxetable}{rcccccccc}
\tablewidth{0pt}
\tablecaption{Observation details, coordinates and radial velocities for our spectroscopic targets.\label{tab:data}}
\tablehead{
  ID       & spectrum&exp time &Obs-date & airmass &RA    &  DEC    &   RV   &RV$_{\rm helio}$    \\    
           &         &  [s]    &         &         &J2000 &J2000    & [\kmsec]& [\kmsec]}
\startdata
NGC6934-n2 & MIKE   &2$\times$1800                 &2017/05/05   & $\sim$1.5 &20:34:10.694  & 07:24:19.393 &$-$433.37 &$-$406.29\\
           & GRACES &4$\times$1050                 &2015/08/13   & $\sim$1.1 &  &  &$-$404.15 &$-$407.93\\
NGC6934-a2 & MIKE   &2$\times$1800                 &2017/05/05   & $\sim$1.3 &20:34:07.579  & 07:24:17.220  &$-$427.70 &$-$400.59\\
           & GRACES &2$\times$1000 - 2$\times$1020 &2015/12/15-16& $\sim$1.9 &  &  &$-$380.23 &$-$401.37\\
NGC6934-n1 & GRACES &4$\times$1010                 &2015-08-13   & $\sim$1.3 &20:34:10.399  &07:24:17.964  &$-$407.95 &$-$411.89\\
NGC6934-a1 & GRACES &2$\times$1550                 &2015-12-18-22& $\sim$1.9 &20:34:16.460  & 07:24:53.374  &$-$382.42 &$-$402.09\\
\enddata
\end{deluxetable}

\movetabledown=6.2cm
\begin{rotatetable}
\floattable

\begin{deluxetable}{cccccccccccccccccccc}
\tablewidth{12pt}
\tablecaption{Atmospheric parameters, derived from spectroscopy and
  photometry, and corresponding Fe~\,{\sc i} and Fe~\,{\sc ii}
  abundances (with the associated $\sigma$ and number of spectral
  lines) for our targets. \\ 
  Results are listed for both MIKE and GRACES spectra.\label{tab:atm}}
\tablehead{
  ID       & SOURCE  &\teff & \logg & [A/H]   & \vmicro &$log~\epsilon$(Fe\,{\sc i})&$\sigma$&\#& $log~\epsilon$(Fe\,{\sc ii})&$\sigma$&\# &\teff & \logg & [A/H] & \vmicro &$log~\epsilon$(Fe\,{\sc i})&$\sigma$ & $log~\epsilon$(Fe\,{\sc ii})&$\sigma$ \\    
           &         &  (K) & (cgs) & (dex)   &(\kmsec) &                           &        &  &                             &        &   &(K)   & (cgs) & (dex) &(\kmsec) &                           &         &                             &         \\
           &         & \multicolumn{10}{c}{Spectroscopic}  &  \multicolumn{8}{c}{Photometric}      
}
\startdata
NGC6934-n2 & MIKE    &4450  & 1.25  & $-$1.53 & 1.78    & 5.968                     & 0.122  & 153  & 6.058                   & 0.097  &17 &4372  & 1.22  &$-$1.61& 1.78    & 5.886                     & 0.126   & 6.105                       & 0.105 \\
           & GRACES  &4500  & 1.10  & $-$1.37 & 1.77    & 6.127                     & 0.142  & 119  & 6.205                   & 0.169  &10 &4372  & 1.22  &$-$1.46& 1.69    & 6.038                     & 0.149   & 6.398                       & 0.172 \\
NGC6934-a2 & MIKE    &4450  & 1.40  & $-$1.35 & 1.80    & 6.143                     & 0.128  & 157  & 6.240                   & 0.133  &18 &4380  & 1.30  &$-$1.39& 1.75    & 6.096                     & 0.128   & 6.282                       & 0.141 \\
           & GRACES  &4550  & 1.60  & $-$1.14 & 2.00    & 6.361                     & 0.196  & 106  & 6.425                   & 0.113  & 8 &4380  & 1.30  &$-$1.27& 1.90    & 6.231                     & 0.204   & 6.494                       & 0.114 \\
NGC6934-n1 & GRACES  &4445  & 1.13  & $-$1.39 & 1.95    & 6.108                     & 0.141  & 112  & 6.188                   & 0.101  &11 &4329  & 1.13  &$-$1.48& 1.88    & 6.019                     & 0.146   & 6.287                       & 0.114 \\
NGC6934-a1 & GRACES  &4650  & 1.50  & $-$1.10 & 1.97    & 6.399                     & 0.165  &  65  & 6.504                   & 0.262  & 4 &4459  & 1.48  &$-$1.25& 1.85    & 6.252                     & 0.178   & 6.648                       & 0.230 \\
\enddata
\end{deluxetable}
\end{rotatetable}

\begin{deluxetable}{ccccrrrrrr}
\tablecaption{Line list with the measured EWs for the program stars observed with MIKE and GRACES. Only a portion of this table is shown here to demonstrate its form and content. A machine-readable version will be available.}\label{tab:lines}
\tablehead{
Wavelength & Species   & L.E.P. & \loggf  & n2           & a2                 & n2       & a2      & n1          & a1\\ 
$[\AA]$  &          & [eV]   &            & \multicolumn{2}{c}{${\rm EWs_{MIKE} [m\AA]}$}&\multicolumn{4}{c}{${\rm EWs_{GRACES}  [m\AA]}$}    }
\startdata
4139.94 & Fe\,{\sc i} & 0.99 & $-$3.63 & 85.7  & 107.2  &--     &-- &--	    & -- \\
4439.89 & Fe\,{\sc i} & 2.28 & $-$3.00 & 69.7  & 70.7   &--     &-- &--	    & -- \\
4445.48 & Fe\,{\sc i} & 0.09 & $-$5.44 & 92.3  & 99.6   &--     &-- &--	    & -- \\
4489.75 & Fe\,{\sc i} & 0.12 & $-$3.97 & 152.4 & 156.4  &--     &-- &--	    & -- \\
4574.23 & Fe\,{\sc i} & 3.21 & $-$2.36 & 34.9  & 37.9   &--     &-- &--     & -- \\
4602.01 & Fe\,{\sc i} & 1.61 & $-$3.15 & 100.2 & 112.0  &90.6   &-- &133.9  & -- \\
4602.95 & Fe\,{\sc i} & 1.49 & $-$2.22 & 148.0 & 164.4  &148.7  &-- &--     & -- \\
4658.30 & Fe\,{\sc i} & 3.27 & $-$2.96 & --    & 18.8   &--     &-- &--	    & -- \\
4788.76 & Fe\,{\sc i} & 3.24 & $-$1.54 & 61.9  & 71.9   &71.7   &-- &--	    & -- \\
4839.55 & Fe\,{\sc i} & 3.27 & $-$1.84 & 61.7  & 73.6   &70.2   &-- &--	    & -- \\
4859.74 & Fe\,{\sc i} & 2.87 & $-$0.78 & 139.0 & 149.6  &--     &-- &--	    & -- \\
4885.43 & Fe\,{\sc i} & 3.88 & $-$1.15 & 58.6  & 61.0   &--     &-- &--	    & -- \\
4917.24 & Fe\,{\sc i} & 4.19 & $-$1.27 & 45.1  & 52.3   &59.1   &-- &41.7   & -- \\
4924.77 & Fe\,{\sc i} & 2.28 & $-$2.29 & 111.2 & 122.8  &114.8  &-- &128.9  & -- \\
\enddata
\end{deluxetable}

\begin{table}
\tiny
\caption{Analysed chemical abundances from C to Ni.\label{tab:abu1}}
\begin{splittabular}{cccccccccccccccccccBcccccccccccccccccBccccccccccccccccccccBc}
  ID       & spectrum&[C/Fe] & $\sigma$ & \# & [O/Fe] &$\sigma$&\#&[Na/Fe]&$\sigma$&[Na/Fe]$_{\rm {NLTE}}$&$\sigma$&\# &[Mg/Fe] & $\sigma$ & \# & [Al/Fe] &$\sigma$&\# &ID       & spectrum&[Si/Fe] & $\sigma$& \#& [Ca/Fe]&$\sigma$&\#&[Sc/Fe]~{\sc ii}&$\sigma$&\#&[Ti/Fe]~{\sc i} & $\sigma$ &\#&[Ti/Fe]~{\sc ii} &$\sigma$&\#&ID       & spectrum&[V/Fe] & $\sigma$& \#& [Cr/Fe]~{\sc i}&$\sigma$&\#&[Cr/Fe]~{\sc ii}&$\sigma$&\#&[Mn/Fe] & $\sigma$ &\#&[Co/Fe] &$\sigma$&\#&[Ni/Fe] &$\sigma$&\# \\\hline
NGC6934-n2 & MIKE    &$-$0.58& 0.02     & 2  &  0.57  & 0.01   &2 & 0.05  & 0.24   &$-$0.13& 0.08   & 5 & 0.39   & 0.00     & 2  & $-$0.20 & 0.13   & 2 &NGC6934-n2 & MIKE    & 0.38 & 0.13  & 9  & 0.32  & 0.10   & 17 & 0.05  & 0.20   & 6  & 0.19 & 0.10 & 29 & 0.35 & 0.16 & 9&  NGC6934-n2 & MIKE    & 0.03 & 0.11  & 14  &$-$0.11   & 0.09   & 10 & 0.14  & --   & 1 &$-$0.46 & 0.13 & 9 &$-$0.13& --   &1 &$-$0.05&0.14&24 & aa\\
           & GRACES  &   --  & --       & -  &  0.50  & 0.08   &2 & 0.08  & 0.31   &$-$0.08& 0.30   & 3 & 0.37   & 0.04     & 2  & $-$0.14 & 0.17   & 2 &           & GRACES  & 0.55 & 0.08  & 4  & 0.33  & 0.09   & 17 & 0.00  & 0.27   & 7  & 0.14 & 0.23 & 21 & 0.32 & 0.34 & 3 &           & GRACES  &-0.07 & 0.14  & 10  &$-$0.03   & 0.17   &  8 &$-$0.06& --   & 1 &$-$0.62 & 0.15 & 9 &  0.02 & --   &1 &$+$0.07&0.24&14 & aa\\
NGC6934-a2 & MIKE    &$-$0.56& 0.01     & 2  &  0.72  & 0.08   &2 & 0.00  & 0.23   &$-$0.17& 0.07   & 6 & 0.36   & 0.00     & 2  & $-$0.15 & 0.05   & 2 &NGC6934-a2 & MIKE    & 0.35 & 0.13  & 11 & 0.32  & 0.10   & 19 & 0.09  & 0.12   & 7  & 0.17 & 0.16 & 32 & 0.33 & 0.12 &10&  NGC6934-a2 & MIKE    & 0.05 & 0.09  & 14  &$-$0.11   & 0.09   & 10 & 0.16  & 0.24 & 2 &$-$0.42 & 0.08 & 9 &$-$0.19& 0.09 &2 &$-$0.01&0.14&23 & aa \\
           & GRACES  &   --  & --       & -  &  0.58  & 0.13   &2 &$-$0.27& 0.28   &$-$0.39& 0.15   & 4 & 0.34   & 0.07     & 2  &  --     & --     & - &           & GRACES  & 0.47 & 0.27  & 2  & 0.35  & 0.24   & 18 &$-$0.13& 0.35   & 4  & 0.21 & 0.16 & 16 & 0.12 & 0.18 & 2  &          & GRACES  & 0.02 & 0.18  & 10  &$-$0.09   & 0.38   &  5 & --    & --   &-- &$-$0.66 & 0.09 & 5 &  --   & --   &--&$-$0.09&0.14&14 & aa  \\
NGC6934-n1 & GRACES  &   --  & --       & -  &  0.64  & 0.04   &2 & 0.22  & 0.08   & 0.09  & 0.09   & 5 & 0.48   & 0.11     & 2  & $+$0.13 & 0.02   & 2 &NGC6934-n1 & GRACES  & 0.50 & 0.04  & 4  & 0.25  & 0.12   & 16 &$-$0.06& 0.12   & 6  & 0.21 & 0.25 & 18 & 0.32 & 0.15 & 4&  NGC6934-n1 & GRACES  &-0.09 & 0.12  &  9  &$-$0.06   & 0.30   &  4 & --    & --   &-- &$-$0.62 & 0.12 & 5 &$-$0.06& --   &1 &$+$0.01&0.13&17 & aa   \\
NGC6934-a1 & GRACES  &   --  & --       & -  &  0.63  & 0.13   &2 & 0.17  & --     &$-$0.18& --     & 1 & 0.54   & --       & 1  & $-$0.08 & --     & 1 &NGC6934-a1 & GRACES  & 0.53 & --    & 1  & 0.33  & 0.20   & 12 & --    & --     & -- & 0.15 & 0.09 & 8  & --   & --   &--&  NGC6934-a1 & GRACES  & 0.10 & 0.14  &  6  &  --      &  --    & -- &   --  & --   & --&$-$0.59 & 0.14 & 5 &$-$0.23& --   &1 &$-$0.01&0.16&7  & aa  \\
\end{splittabular}
\end{table}

\begin{table}
\scriptsize
\caption{Analysed chemical abundances from Cu to Dy.\label{tab:abu2}}
\begin{splittabular}{cc ccc ccc ccc ccc cccBcc ccc ccc ccc ccc Bcc ccc ccc ccc ccc}
  ID       & spectrum&[Cu/Fe] & $\sigma$& \#& [Zn/Fe]&$\sigma$&\#&[Sr/Fe]~{\sc ii}&$\sigma$&\#&[Y/Fe]~{\sc ii} & $\sigma$ &\#&[Zr/Fe]~{\sc ii} &$\sigma$&\#&ID&spectrum&[Ba/Fe]~{\sc ii}&$\sigma$&\# &ID       & spectrum&[La/Fe]~{\sc ii}&$\sigma$&\#&[Ce/Fe]~{\sc ii}&$\sigma$&\#&[Pr/Fe]~{\sc ii} & $\sigma$ &\#&[Nd/Fe]~{\sc ii} &$\sigma$&\#&[Sm/Fe]~{\sc ii} &$\sigma$&\#&[Eu/Fe]~{\sc ii} &$\sigma$&\#&[Dy/Fe]~{\sc ii} &$\sigma$&\# \\\hline
NGC6934-n2 & MIKE   &$-$0.59 & 0.16 & 2 &0.15&0.01&2 & 0.55  & -- &1 &-0.01& 0.02 & 4 &0.24&--  &1 &NGC6934-n2 & MIKE   &0.09&0.09&3   &0.26&0.09& 8 &0.22&0.07& 7&0.31&0.03&2 &NGC6934-n2 & MIKE    &0.34&0.09 & 18&0.50&0.07&14&0.63& -- &1&0.83&--&1 \\
           & GRACES &$-$0.75 & 0.15 & 2 & -- & -- &--& --    & -- &--&-0.07& 0.17 & 4 & -- &--  &--&           & GRACES &0.10&0.05&3   &0.27&0.13& 6 &0.15& -- & 1&0.34& -- &1 &           & GRACES  &0.36&0.09 & 10&0.59&0.17& 3&0.50& -- &1&--&--&--   \\
NGC6934-a2 & MIKE   &$-$0.66 & 0.01 & 2 &0.14&0.11&2 & 0.60  & -- &1 & 0.01& 0.06 & 4 &0.21&--  &1 &NGC6934-a2 & MIKE   &0.04&0.05&3   &0.31&0.08& 8 &0.19&0.09& 7&0.31&0.13&2 &NGC6934-a2 & MIKE    &0.33&0.14 & 18&0.50&0.13&14&0.53& -- &1&0.82&--&1  \\
           & GRACES &$-$0.73 & --   & 1 &--  &--  &--& --    & -- &--& --  & --   & --& -- &--  &--&           & GRACES &$-$0.01&0.06&3&0.20&0.05& 2 & -- & -- &--& -- &--  &--&           & GRACES  &0.49&0.05&  3& -- & -- &--&0.63& -- &1&--&--&--   \\
NGC6934-n1 & GRACES &$-$0.80 & --   & 1 &--  &--  &--& --    & -- &--&-0.06& 0.10 & 2 & -- &--  &--&NGC6934-n1 & GRACES &0.10&0.08&3   &0.23&0.12& 6 &0.19&0.03& 2&0.21&--  & 1&NGC6934-n1 & GRACES  &0.31&0.11&11 &0.57&0.25& 4&0.50& -- &1&--&--&--   \\
NGC6934-a1 & GRACES &$-$0.75 & --   & 1 & -- &--  &--& --    & -- &--& --  & --   & --& -- &--  &--&NGC6934-a1 & GRACES &0.17&0.23&3   &0.10& -- & 1 & -- & -- &--& -- &--  &--&NGC6934-a1 & GRACES  &0.36&0.04&  3& -- &--  &--&0.55& -- &1&--&--&--    \\
\end{splittabular}
\end{table}


\startlongtable
\begin{deluxetable}{lcccccc}
\tablewidth{10pt}
\tablecaption{Sensitivity of the derived abundances to the uncertainties in atmospheric parameters and uncertainties due to the errors in the EWs measurements or in the $\chi$-square fitting procedure. We reported the total internal uncertainty ($\sigma_{\rm total}$) obtained by the quadratic sum of all the contributers to the error.}\label{tab:err}
\tablehead{
      &\colhead{$\Delta$\teff} &\colhead{$\Delta$\logg}&\colhead{$\Delta$\vmicro}&\colhead{$\Delta$[A/H]}& \colhead{$\sigma_{\rm EWs/fit}$}&\colhead{$\sigma_{\rm total}$}\\  
      &\colhead{$\pm$50~K}    & \colhead{$\pm$0.20}   & \colhead{$\pm$0.20~\kmsec} & \colhead{$\pm$0.10~dex} &     &                   }
\startdata
\multicolumn{7}{c}{MIKE}\\
$\rm {[C/Fe]}$           & $\pm$0.07   & $\pm$0.00  & $\pm$0.00  & $\mp$0.03 & $\pm$0.06           & $\pm$0.10 \\
$\rm {[O/Fe]}$           & $\pm$0.00   & $\pm$0.07  & $\mp$0.00  & $\pm$0.03 & $\pm$0.09           & $\pm$0.12  \\
$\rm {[Na/Fe]}$          & $\mp$0.01   & $\mp$0.01  & $\pm$0.03  & $\mp$0.01 & $\pm$0.04           & $\pm$0.05 \\
$\rm {[Mg/Fe]}$          & $\mp$0.01   & $\mp$0.02  & $\mp$0.01  & $\pm$0.00 & $\pm$0.07           & $\pm$0.07   \\
$\rm {[Al/Fe]}$          & $\pm$0.05   & $\mp$0.00  & $\pm$0.01  & $\pm$0.00 & $\pm$0.14           & $\pm$0.15 \\
$\rm {[Si/Fe]}$          & $\mp$0.07   & $\pm$0.01  & $\pm$0.06  & $\pm$0.02 & $\pm$0.03           & $\pm$0.10    \\
$\rm {[Ca/Fe]}$          & $\pm$0.00   & $\mp$0.01  & $\mp$0.01  & $\mp$0.01 & $\pm$0.02           & $\pm$0.03 \\
$\rm {[Sc/Fe]}$\,{\sc ii}& $\pm$0.05   & $\mp$0.00  & $\pm$0.02  & $\mp$0.00 & $\pm$0.04           & $\pm$0.07    \\
$\rm {[Ti/Fe]}$\,{\sc i} & $\pm$0.04   & $\mp$0.01  & $\pm$0.01  & $\mp$0.02 & $\pm$0.02           & $\pm$0.05    \\
$\rm {[Ti/Fe]}$\,{\sc ii}& $\pm$0.04   & $\mp$0.01  & $\mp$0.04  & $\mp$0.01 & $\pm$0.03           & $\pm$0.07    \\
$\rm {[V/Fe]}$           & $\pm$0.05   & $\mp$0.00  & $\pm$0.04  & $\mp$0.01 & $\pm$0.02           & $\pm$0.07    \\
$\rm {[Cr/Fe]}$\,{\sc i} & $\pm$0.03   & $\mp$0.01  & $\mp$0.00  & $\mp$0.01 & $\pm$0.03           & $\pm$0.04    \\
$\rm {[Cr/Fe]}$\,{\sc ii}& $\pm$0.00   & $\mp$0.00  & $\pm$0.02  & $\mp$0.01 & $\pm$0.10           & $\pm$0.10    \\
$\rm {[Mn/Fe]}$          & $\pm$0.08   & $\pm$0.00  & $\mp$0.02  & $\mp$0.02 & $\pm$0.05           & $\pm$0.10    \\
$\rm {[Fe/H]}$\,{\sc i}  & $\pm$0.06   & $\pm$0.00  & $\mp$0.07  & $\mp$0.00 & $\pm$0.01           & $\pm$0.09   \\
$\rm {[Fe/H]}$\,{\sc ii} & $\mp$0.05   & $\pm$0.07  & $\mp$0.06  & $\pm$0.04 & $\pm$0.03           & $\pm$0.11     \\
$\rm {[Co/Fe]}$          & $\pm$0.00   & $\pm$0.00  & $\pm$0.06  & $\pm$0.01 & $\pm$0.10           & $\pm$0.12    \\
$\rm {[Ni/Fe]}$          & $\mp$0.01   & $\pm$0.01  & $\pm$0.04  & $\pm$0.01 & $\pm$0.02           & $\pm$0.05    \\
$\rm {[Cu/Fe]}$          & $\pm$0.08   & $\mp$0.00  & $\mp$0.07  & $\pm$0.00 & $\mp$0.06           & $\pm$0.12    \\
$\rm {[Zn/Fe]}$          & $\mp$0.10   & $\pm$0.03  & $\pm$0.01  & $\pm$0.03 & $\pm$0.08           & $\pm$0.14    \\
$\rm {[Sr/Fe]}$\,{\sc ii}& $\pm$0.09   & $\mp$0.01  & $\mp$0.06  & $\mp$0.03 & $\pm$0.08           & $\pm$0.14    \\
$\rm {[Y/Fe]}$\,{\sc ii}  & $\pm$0.00  & $\pm$0.05  & $\mp$0.09  & $\pm$0.00 & $\pm$0.04           & $\pm$0.11    \\
$\rm {[Zr/Fe]}$\,{\sc ii} & $\pm$0.02  & $\pm$0.06  & $\mp$0.02  & $\pm$0.01 & $\pm$0.06           & $\pm$0.09    \\
$\rm {[Ba/Fe]}$\,{\sc ii} & $\pm$0.01  & $\pm$0.07  & $\mp$0.13  & $\pm$0.03 & $\pm$0.04           & $\pm$0.16    \\
$\rm {[La/Fe]}$\,{\sc ii}& $\pm$0.01   & $\pm$0.07  & $\mp$0.02  & $\pm$0.03 & $\pm$0.03           & $\pm$0.08    \\
$\rm {[Ce/Fe]}$\,{\sc ii}& $\pm$0.00   & $\pm$0.05  & $\mp$0.05  & $\pm$0.02 & $\pm$0.04           & $\pm$0.08    \\
$\rm {[Pr/Fe]}$\,{\sc ii}& $\pm$0.00   & $\pm$0.06  & $\pm$0.00  & $\pm$0.04 & $\pm$0.09           & $\pm$0.12    \\
$\rm {[Nd/Fe]}$\,{\sc ii}& $\pm$0.02   & $\pm$0.06  & $\mp$0.04  & $\pm$0.01 & $\pm$0.02           & $\pm$0.08    \\
$\rm {[Sm/Fe]}$\,{\sc ii}& $\pm$0.01   & $\pm$0.06  & $\mp$0.03  & $\pm$0.01 & $\pm$0.02           & $\pm$0.07    \\
$\rm {[Eu/Fe]}$\,{\sc ii}& $\mp$0.01   & $\pm$0.07  & $\pm$0.00  & $\pm$0.04 & $\pm$0.09           & $\pm$0.12    \\
$\rm {[Dy/Fe]}$\,{\sc ii}& $\pm$0.03   & $\pm$0.08  & $\mp$0.06  & $\mp$0.01 & $\pm$0.08           & $\pm$0.13    \\
\hline
\multicolumn{7}{c}{GRACES {\it normal} stars}\\
$\rm {[O/Fe]}$           & $\pm$0.01   & $\pm$0.07  & $\pm$0.00  & $\pm$0.04 & $\pm$0.14           & $\pm$0.16  \\
$\rm {[Na/Fe]}$          & $\mp$0.01   & $\mp$0.02  & $\pm$0.04  & $\mp$0.01 & $\pm$0.05           & $\pm$0.07 \\
$\rm {[Mg/Fe]}$          & $\mp$0.01   & $\mp$0.03  & $\pm$0.01  & $\mp$0.00 & $\pm$0.08           & $\pm$0.09   \\
$\rm {[Al/Fe]}$          & $\pm$0.03   & $\mp$0.00  & $\pm$0.00  & $\pm$0.00 & $\pm$0.18           & $\pm$0.18 \\
$\rm {[Si/Fe]}$          & $\mp$0.09   & $\pm$0.01  & $\pm$0.06  & $\pm$0.02 & $\pm$0.06           & $\pm$0.13    \\
$\rm {[Ca/Fe]}$          & $\pm$0.00   & $\mp$0.02  & $\mp$0.01  & $\mp$0.01 & $\pm$0.02           & $\pm$0.03 \\
$\rm {[Sc/Fe]}$\,{\sc ii}& $\pm$0.01   & $\mp$0.00  & $\pm$0.00  & $\mp$0.02 & $\pm$0.04           & $\pm$0.05    \\
$\rm {[Ti/Fe]}$\,{\sc i} & $\pm$0.05   & $\mp$0.01  & $\pm$0.00  & $\mp$0.02 & $\pm$0.02           & $\pm$0.06    \\
$\rm {[Ti/Fe]}$\,{\sc ii}& $\pm$0.01   & $\mp$0.01  & $\mp$0.06  & $\mp$0.02 & $\pm$0.06           & $\pm$0.09    \\
$\rm {[V/Fe]}$           & $\pm$0.07   & $\mp$0.00  & $\pm$0.07  & $\mp$0.01 & $\pm$0.03           & $\pm$0.10    \\
$\rm {[Cr/Fe]}$\,{\sc i} & $\pm$0.05   & $\mp$0.01  & $\mp$0.04  & $\mp$0.02 & $\pm$0.04           & $\pm$0.08    \\
$\rm {[Cr/Fe]}$\,{\sc ii}& $\mp$0.02   & $\mp$0.00  & $\pm$0.03  & $\mp$0.02 & $\pm$0.11           & $\pm$0.12    \\
$\rm {[Mn/Fe]}$          & $\pm$0.09   & $\mp$0.00  & $\mp$0.02  & $\mp$0.02 & $\pm$0.05           & $\pm$0.11    \\
$\rm {[Fe/H]}$\,{\sc i}  & $\pm$0.06   & $\pm$0.00  & $\mp$0.09  & $\pm$0.00 & $\pm$0.01           & $\pm$0.11   \\
$\rm {[Fe/H]}$\,{\sc ii} & $\mp$0.04   & $\pm$0.06  & $\mp$0.07  & $\pm$0.05 & $\pm$0.04           & $\pm$0.12     \\
$\rm {[Co/Fe]}$          & $\pm$0.01   & $\pm$0.00  & $\pm$0.07  & $\pm$0.01 & $\pm$0.09           & $\pm$0.11    \\
$\rm {[Ni/Fe]}$          & $\mp$0.01   & $\pm$0.01  & $\pm$0.03  & $\pm$0.01 & $\pm$0.03           & $\pm$0.05    \\
$\rm {[Cu/Fe]}$          & $\pm$0.10   & $\pm$0.00  & $\mp$0.07  & $\mp$0.02 & $\pm$0.09           & $\pm$0.15    \\
$\rm {[Y/Fe]}$\,{\sc ii} & $\pm$0.02   & $\pm$0.04  & $\mp$0.08  & $\mp$0.01 & $\pm$0.11           & $\pm$0.14    \\
$\rm {[Ba/Fe]}$\,{\sc ii}& $\pm$0.01   & $\pm$0.07  & $\mp$0.15  & $\pm$0.04 & $\pm$0.07           & $\pm$0.18    \\
$\rm {[La/Fe]}$\,{\sc ii}& $\pm$0.01   & $\pm$0.06  & $\mp$0.01  & $\pm$0.04 & $\pm$0.07           & $\pm$0.10    \\
$\rm {[Ce/Fe]}$\,{\sc ii}& $\mp$0.02   & $\pm$0.04  & $\mp$0.06  & $\pm$0.02 & $\pm$0.20           & $\pm$0.21    \\
$\rm {[Pr/Fe]}$\,{\sc ii}& $\pm$0.00   & $\pm$0.08  & $\pm$0.00  & $\pm$0.04 & $\pm$0.20           & $\pm$0.22    \\
$\rm {[Nd/Fe]}$\,{\sc ii}& $\pm$0.00   & $\pm$0.06  & $\mp$0.04  & $\pm$0.02 & $\pm$0.06           & $\pm$0.10    \\
$\rm {[Sm/Fe]}$\,{\sc ii}& $\mp$0.01   & $\pm$0.03  & $\mp$0.05  & $\pm$0.00 & $\pm$0.20           & $\pm$0.21    \\
$\rm {[Eu/Fe]}$\,{\sc ii}& $\pm$0.00   & $\pm$0.08  & $\pm$0.00  & $\pm$0.05 & $\pm$0.10           & $\pm$0.14    \\
\hline
\multicolumn{7}{c}{GRACES {\it anomalous} stars}\\
$\rm {[O/Fe]}$           & $\pm$0.00   & $\pm$0.06  & $\pm$0.00  & $\pm$0.05 & $\pm$0.20           & $\pm$0.21  \\
$\rm {[Na/Fe]}$          & $\mp$0.00   & $\mp$0.01  & $\pm$0.06  & $\mp$0.01 & $\pm$0.08           & $\pm$0.10 \\
$\rm {[Mg/Fe]}$          & $\mp$0.01   & $\mp$0.03  & $\pm$0.00  & $\mp$0.01 & $\pm$0.09           & $\pm$0.10   \\
$\rm {[Al/Fe]}$          & $\pm$0.03   & $\mp$0.00  & $\pm$0.01  & $\mp$0.01 & $\pm$0.22           & $\pm$0.22 \\
$\rm {[Si/Fe]}$          & $\mp$0.08   & $\pm$0.01  & $\pm$0.06  & $\pm$0.02 & $\pm$0.12           & $\pm$0.16    \\
$\rm {[Ca/Fe]}$          & $\pm$0.02   & $\mp$0.01  & $\mp$0.01  & $\mp$0.02 & $\pm$0.03           & $\pm$0.04 \\
$\rm {[Sc/Fe]}$\,{\sc ii}& $\pm$0.05   & $\mp$0.00  & $\pm$0.00  & $\mp$0.01 & $\pm$0.07           & $\pm$0.09    \\
$\rm {[Ti/Fe]}$\,{\sc i} & $\pm$0.06   & $\pm$0.00  & $\pm$0.03  & $\mp$0.02 & $\pm$0.03           & $\pm$0.08    \\
$\rm {[Ti/Fe]}$\,{\sc ii}& $\pm$0.06   & $\mp$0.00  & $\mp$0.05  & $\mp$0.01 & $\pm$0.06           & $\pm$0.10    \\
$\rm {[V/Fe]}$           & $\pm$0.06   & $\mp$0.00  & $\pm$0.05  & $\mp$0.02 & $\pm$0.03           & $\pm$0.09    \\
$\rm {[Cr/Fe]}$\,{\sc i} & $\pm$0.05   & $\mp$0.00  & $\pm$0.00  & $\mp$0.02 & $\pm$0.06           & $\pm$0.08    \\
$\rm {[Mn/Fe]}$          & $\pm$0.10   & $\pm$0.00  & $\mp$0.01  & $\mp$0.02 & $\pm$0.06           & $\pm$0.12    \\
$\rm {[Fe/H]}$\,{\sc i}  & $\pm$0.05   & $\pm$0.00  & $\mp$0.09  & $\pm$0.01 & $\pm$0.01           & $\pm$0.10   \\
$\rm {[Fe/H]}$\,{\sc ii} & $\mp$0.08   & $\pm$0.06  & $\mp$0.06  & $\pm$0.04 & $\pm$0.05           & $\pm$0.13     \\
$\rm {[Co/Fe]}$          & $\mp$0.05   & $\mp$0.00  & $\pm$0.09  & $\mp$0.01 & $\pm$0.20           & $\pm$0.22    \\
$\rm {[Ni/Fe]}$          & $\mp$0.00   & $\pm$0.01  & $\pm$0.04  & $\pm$0.02 & $\pm$0.03           & $\pm$0.05    \\
$\rm {[Cu/Fe]}$          & $\pm$0.05   & $\pm$0.01  & $\mp$0.01  & $\mp$0.01 & $\pm$0.12           & $\pm$0.13    \\
$\rm {[Ba/Fe]}$\,{\sc ii}& $\pm$0.01   & $\pm$0.06  & $\mp$0.15  & $\pm$0.03 & $\pm$0.10           & $\pm$0.19    \\
$\rm {[La/Fe]}$\,{\sc ii}& $\pm$0.01   & $\pm$0.07  & $\pm$0.00  & $\pm$0.04 & $\pm$0.12           & $\pm$0.14    \\
$\rm {[Nd/Fe]}$\,{\sc ii}& $\pm$0.01   & $\pm$0.06  & $\mp$0.06  & $\pm$0.04 & $\pm$0.11           & $\pm$0.14    \\
$\rm {[Eu/Fe]}$\,{\sc ii}& $\mp$0.02   & $\pm$0.05  & $\pm$0.00  & $\pm$0.02 & $\pm$0.13           & $\pm$0.14    \\
\enddata
\end{deluxetable}


%
\movetabledown=6.2cm
\begin{rotatetable}
\tiny
\begin{deluxetable}{l | c l c c l c c l c c l | c}
\tablecaption{List of GCs with confirmed chemical and/or photometric anomalies. 
The listed chemical properties, as measured from spectroscopic data, 
include: {\it (i)} star-to-star variations in Fe, \\{\it (ii)} $s$-process
elements, {\it (iii)} multiple $p$-capture (anti-)correlation (e.g. within groups
of stars with different Fe and $s$-elements as displayed in Fig.~14 in
Marino et al.\,2011), {\it (iv)} differences in \\ the overall C+N+O. 
For each GC we list a ``Proposed Class'', which refers to the presence
of variations in Fe (Iron-II), in $s$-elements ($s$-II), and multiple
sequence on the chromosome map, \\ as found in Milone et al.\,(2017,
Type~II). Further details in Section~\ref{sec:conclusions}.}\label{tab:anomaliclass}
\tablehead{
\colhead{GC\phantom{1111111111111}}                           & \multicolumn{11}{c}{CHEMICAL ABUNDANCE VARIATIONS} & \colhead{Proposed Class}
 }
\startdata
                             &metallicity   &Literature        &&$s$-elements&Literature               &&$p$-capture elements&Literature             && C+N+O & Literature    &                \\
                             &              &                        &&            &                                &&  in each Fe group  &                              &&       &                      &                \\\hline
NGC\,1261                    & (?)          &                        &&(?)         &                                && (?)                &                              &&(?)    &                      &Type~II         \\\hline
NGC\,1851                    &possible small&Carretta+10 &&yes         &YG08\tablenotemark{a};       && yes                &Carretta+10;      &&yes    &Yong+14   &Iron-II/$s$-II/Type~II\\
                             &              &Gratton+13  &&            &Villanova+10        &&                    &Villanova+10      &&       &                      &                \\
                             &              &Marino+14   &&            &                                &&                    &                              &&       &                      &                \\\hline
NGC\,362                     & (?)          &                        &&yes         &   Carretta+13      && (?)                &                              &&(?)    &                      &$s$-II/Type~II         \\\hline
NGC\,5139 &yes           &Norris+96;  &&yes         &NDaC95\tablenotemark{b};     && yes                &JP10\tablenotemark{c};&&yes   & Marino+12b&Iron-II/$s$-II/Type~II\\
($\omega$~Centauri)                             &              &SK96\tablenotemark{d}        &&           &Smith+00;           &&                    &Marino+11b        &&       &                      &                \\
                             &              &                        &&            &JP10\tablenotemark{c}; &&                    &                              &&       &                      &                \\
                             &              &                        &&            &Marino+11b;         &&                    &                              &&       &                      &                \\
                             &              &                        &&            &D'Orazi+11          &&                    &                              &&       &                      &                \\\hline
NGC\,5286                    &yes           &Marino+15   &&yes         &Marino+15           && yes                &Marino+15         &&not-studied&                  &Iron-II/$s$-II/Type~II\\\hline
NGC\,5824                    &yes(?)        &Da Costa+14 &&yes(?)      &Roederer+16         && not-studied        &                              &&not-studied&                  &Iron-II(?)/$s$-II(?)\\\hline
NGC\,6229                    &possible small&Johnson+17  &&yes         &Johnson+17          && no(?)              &Johnson+17        &&not-studied&                  &Iron-II(?)/$s$-II\\\hline
NGC\,6273 (M\,19)            &yes           &Johnson+15  &&yes         &Johnson+15          && yes                &Johnson+15        &&not-studied&                  &Iron-II/$s$-II   \\\hline
NGC\,6388                    & (?)          &                        &&(?)         &                                && (?)                &                              &&(?)    &                      &Type~II         \\\hline
NGC\,6656 (M\,22)            &yes           &Marino+09;  &&yes         &Marino+09,11a,12a&& yes             &Marino+09,11a  &&yes    &Marino+11a &Iron-II/$s$-II/Type~II\\
                             &              &Da Costa+09 &&            &                                &&                    &                              &&       &Alves Brito+12&             \\\hline
NGC\,6715 (M\,54)            &yes           &Carretta+10 &&yes(?) &BWG99\tablenotemark{e}                                && yes                &Carretta+10       &&not-studied&                  &Iron~II/$s$-II(?)/Type~II \\ 
                             &              &            &&       &                                &&                    &                              &&       &                      &                \\\hline
NGC\,6934                    &yes           &This work               &&no          &This work                       && not-studied        &                              &&not-studied&                  &Iron~II/$s$-I/Type~II \\
                             &              &                        &&            &                                &&                    &                              &&       &                      &                \\\hline
NGC\,7089 (M\,2)             &yes           &Yong+14     &&yes         &Lardo+13;           && yes                &Yong+14           &&not-studied&                  &Iron-II/$s$-II/Type~II\\
                             &              &            &&            &Yong+14             &&                    &                              &&           &                  &                \\\hline
Terzan~5                     &yes           &Ferraro+09  &&not studied &                                && not studied        &                              &&not-studied&                  &Iron-II         \\
                             &              &Origlia+11  &&            &                                &&                    &                              &&       &                      &                \\
                             &              &Massari+14  &&            &                                &&                    &                              &&       &                      &                \\
\enddata
\tablenotetext{a}{Yong \& Grundahl\,(2008)}
\tablenotetext{b}{Norris \& Da Costa\,(1995)}
\tablenotetext{c}{Johnson \& Pilachowski\,(2010)}
\tablenotetext{d}{Suntzeff \& Kraft\,(1996)}
\tablenotetext{e}{Brown, Wallerstein \& Gonzalez\,(1999)}
\end{deluxetable}
\end{rotatetable}

\end{document}